\documentclass[Phys]{SciPost}

\usepackage[utf8]{inputenc}
\usepackage{amsmath}
\usepackage[T1]{fontenc}
\usepackage{amsmath,amsfonts,amssymb,amsthm,epsfig,epstopdf,url,array}
\usepackage{bm}
\usepackage{graphicx}
\usepackage{mathtools}
\usepackage{comment}
\usepackage{color}

\usepackage{dcolumn}
\usepackage{bm}
\usepackage{graphicx}
\usepackage{comment}

\newcommand{\bea}{\begin{eqnarray}}

\newcommand{\eea}{\end{eqnarray}}

\usepackage{color}

\let\oldref\ref
\renewcommand{\ref}[1]{(\oldref{#1})}

\usepackage[utf8]{inputenc}
\usepackage{amsmath}
\usepackage[T1]{fontenc}
\usepackage{amsmath,amsfonts,amssymb,amsthm,epsfig,epstopdf,url,array}
\usepackage{bm}
\usepackage{graphicx}
\usepackage{mathtools}
\usepackage{comment}
\usepackage{color}

\usepackage{dcolumn}
\usepackage{bm}
\usepackage{graphicx}
\usepackage{comment}

\usepackage{color}

\let\oldref\ref
\renewcommand{\ref}[1]{(\oldref{#1})}

\usepackage[utf8]{inputenc}      
\usepackage[T1]{fontenc}         
\usepackage[english]{babel}      
\usepackage{amsmath,amssymb}     
\usepackage{graphicx}            
\usepackage{float}               
\usepackage{booktabs}            

\newcommand{\mk}[1]{\textcolor{black}{#1}}

\begin{document}

\begin{center}
{\Large \textbf{
The distribution of the moment of inertia for harmonically trapped noninteracting Bosons at finite temperature: large deviations 
}}
\end{center}

\begin{center}
Manas Kulkarni\textsuperscript{1}${}^\star$,
Satya N. Majumdar\textsuperscript{2}${}^\dagger$ and
Gr\'egory Schehr\textsuperscript{3}${}^\ddagger$
\end{center}
\begin{center}
{\bf 1}
International Centre for Theoretical Sciences, Tata Institute of Fundamental Research, Bangalore 560089, India \\
\medskip
{\bf 2}
LPTMS, CNRS, Univ.  Paris-Sud,  Universit\'e Paris-Saclay,  91405 Orsay,  France \\
\medskip
{\bf 3}
Sorbonne Universit\'e, Laboratoire de Physique Th\'eorique et Hautes Energies, CNRS UMR 7589, 4 Place Jussieu, 75252 Paris Cedex 05, France

\medskip

${}^\star$ \!{\small \sf manas.kulkarni@icts.res.in} \
${}^\dagger$\! {\small \sf satyanarayan.majumdar@cnrs.fr} \
${}^\ddagger$\! {\small \sf gregory.schehr@lpthe.jussieu.fr}
\end{center}


\section*{Abstract} {\bf We compute the full probability distribution of the moment of inertia $I \propto 
\sum_{i=1}^N \vec{r}_i^{\,2}$ of a gas of $N$ noninteracting bosons trapped in a harmonic potential $V(r) = 
(1/2)\, m\, \omega^2 r^2$, in all dimensions and at all temperature. 
The appropriate thermodynamic limit in a trapped Bose gas consists in taking the limit
$N\to \infty$ and $\omega\to 0$ with their product $\rho = N \omega^d$ fixed, where $\rho$
plays the role analogous to the density in a translationally invariant system. In this thermodynamic limit
and in dimensions $d>1$, the harmonically trapped Bose gas undergoes 
a Bose-Einstein condensation (BEC) transition as the density $\rho$ crosses a 
critical value $\rho_c(\beta)$, where $\beta$ denotes the inverse temperature. \mk{We show that the nature of the condensation
is different for $1<d\leq 2$ and $d>2$. Near the condensation transition, for simplicity, we provide the analysis only for $d>2$.}
We show that the probability 
distribution $P_\beta(I,N)$ of $I$ admits a large deviation form $P_\beta(I,N) \sim e^{-V \Phi(I/V)}$ where $V 
= \omega^{-d} \gg 1$. We compute explicitly the rate function $\Phi(z)$ and show that it exhibits a singularity 
at a critical value $z=z_c$ where its second derivative undergoes a discontinuous jump. In addition, on the condensed side, the rate function $\Phi(z)$ becomes independent of $\rho$ for $z<z_c$. \mk{We show that the existence of $z_c$ and the associated freezing below $z_c$ is directly related to the existence of a BEC transition 
and it disappears when the system does not have a BEC transition as in $d \leq 1$.} An interesting consequence 
of our results is that even if the actual system is in the fluid phase, i.e., when $\rho < \rho_c(\beta)$, by 
measuring the distribution of $I$ and analysing the singularity in the associated rate function, one can get a 
signal of the BEC transition in $d>1$. This provides a real space diagnostic for the BEC transition in
the noninteracting Bose gas.
}

\vspace{10pt}
\noindent\rule{\textwidth}{1pt}
\tableofcontents\thispagestyle{fancy}
\noindent\rule{\textwidth}{1pt}
\vspace{10pt}

\section{Introduction}\label{sec:introduction}

Recent advances in the experimental techniques in cold atom systems have made it possible to explore the 
spatial distribution of quantum particles, bosons and fermions, in the presence of a confining trap
with or without interactions between the particles \cite{BDZ08,NNCS10,Fermicro1,Fermicro2,Fermicro3,Zwi16,Cornell95,Ketterle95,Bradley95,BEC_review02,Leggett01,Stringari99,Yefsah25a,Yefsah25b,Zwier25,Dix25}. The presence of a trap breaks the translational 
symmetry and makes the system inhomogeneous in space.  
In the absence of interactions, the energy spectrum of such systems is trivial and consist of filling up
the single particle energy levels with the appropriate Bose or Fermi statistics. In other words, the many-body
Hamiltonian is trivially quadratic in the second quantised formalism. However, the spatial distributions of such quantum
particles is still nontrivial in such noninteracting systems, due to the fact that the many-body wave functions 
in real space can be rather complicated due to the statistics of the particles. For instance, for fermions, the 
Pauli exclusion principle forbids two fermions (with the same spin) to be in the same position in space or equivalently,
any many-body eigenfunction must have a Slater determinant form, 
which ensures antisymmetry under the exchange
of a pair of fermions. Similarly, for bosons, the many-body eigenfunction must have a permanental form, 
ensuring the symmetry
under the exchange of any pair of bosons. It is also interesting to explore the dependence of these spatial 
properties
on the temperature. For instance for fermions, physical observables exhibit ``quantum behaviors'' when the temperature
is much less than the Fermi energy, while at a higher temperature, they behave more or less classically. {For bosonic systems, on the
other hand, there is a Bose-Einstein condensation (BEC) at a critical temperature $T_c>0$ 
in dimensions $d>d_c$ (where $d_c$ denotes the lower critical dimension that depends on the geometry of the 
confining potential). For example, for a hard-box potential $d_c = 2$, while for a harmonic trap 
$d_c = 1$~\cite{Stringari99}.} This means
that a macroscopically large fraction of bosons condense in the single-particle ground state for 
$T<T_c$ (equivalently when
the density exceeds a critical value $\rho_c$). {
While the mechanism behind the BEC transition for noninteracting bosons is very simple in the energy space, 
it is not completely evident how, by inspecting a configuration of bosons in real space, one can get a 
marked signal for the existence or not of the thermodynamic BEC transition. 
The results of most experimental measurements are reported in momentum space and, 
in fact, there have been relatively few experiments reporting the manifestation of 
BEC in real space~\cite{GWRLO2008, MYK2021}. }

In most of the experimental systems, the number of bosons $N$ is conserved. Hence the natural ensemble to study 
is the canonical one, where one is interested in the canonical partition function $Z(N,\beta)$ where $\beta$ is 
the inverse temperature. However, computing $Z(N,\beta)$ and extracting its large $N$ behavior is somewhat 
cumbersome even for noninteracting bosons and, hence, in most textbooks (see 
e.g.~\cite{Ziff77,Huang_book,Pathria_book}), one usually studies the grand-canonical ensemble where the number of 
particles is allowed to fluctuate around its average value. Theoretically it is easier to study the 
grand-canonical partition function, which is simply the generating function of the canonical partition function 
$Z(N,\beta)$. However, extracting the precise asymptotic behaviours of $Z(N,\beta)$ for large $N$ from its 
generating function is not that straightforward, especially when the system undergoes a BEC phase transition. 
Moreover, even if one could extract the large $N$ behavior of $Z(N,\beta)$ explicitly, it does not directly provide 
any information about the spatial configurations of bosons that we are interested in. This is due to the fact 
that $Z(N,\beta)$, for noninteracting bosons, depends only on the single-particle energies $\epsilon_i$'s and, 
hence, does not carry any information on the spatial properties of the systems, since it does not involve the 
many-body eigenfunctions that encode the spatial distribution.

There have been recent progresses in characterising the spatial properties of noninteracting fermions in a trap 
(harmonic or otherwise), both at zero and finite temperature and several spatial observables have been computed 
explicitly. This includes the probability distribution induced by quantum and thermal fluctuations of the 
position of the farthest fermion from the center of the trap, the full counting statistics, etc 
\cite{fermions_PRA16,fermions_farthest,fermions_rotating,Gouraud} (see e.g. \cite{fermions_JPA19} for a 
review). In this paper, we are interested in a particular spatial observable, namely the distribution of the 
moment of inertia of the trapped noninteracting particles at arbitrary temperature and in arbitrary dimension. 
The moment of inertia $I$ is simply defined as the linear statistics
\begin{equation}
\label{eq:MI}
I \propto  \sum_{i=1}^N \vec{r}_i^{\,2}\, ,
\end{equation}
where ${\vec r}_i$ denotes the position of the $i$-th particle in $d$ dimensions. For noninteracting fermions 
in a harmonic trap, and only in $d=1$, the distribution of $I$ was exactly computed at all temperatures 
\cite{Grela_PRL} and it displays nontrivial large deviation behaviors with a rate function that depends 
explicitly on temperature and displays quite different behaviors in the quantum (where temperature $T$ is of the 
order of the gap in the single particle energy spectrum) and the classical regime (where the temperature is of 
the order of the Fermi energy). It is then natural to ask how this distribution of $I$ behaves for 
noninteracting Bosons in a trap. In particular, for $d>d_c$, where the system undergoes a BEC transition, how 
does this transition affect the statistics of $I$?

In this paper, we address this question and compute exactly the probability distribution of $I$ given in 
Eq.~\eqref{eq:MI} for noninteracting bosons in a harmonic trap in all dimensions and all temperature. This 
exact computation is possible by using an interesting identity that relates the cumulant generating function of 
$I$ with the partition function $Z(N,\beta)$. This identity was derived and used before to compute the finite 
temperature properties of the distribution of $I$ for noninteracting fermions in a harmonic trap in one 
dimension~\cite{Grela_PRL}. Here we show that this identity is more general and can be applied for bosons 
as well.
Moreover, we show that this identity holds even in higher dimensions and at any temperature.  Consequently, it 
gives us access to the cumulant generating function of $I$ and, in turn, to its full distribution at any 
temperature and in any dimension. This observable is a natural one that provides some spatial information about the 
positions of the bosons and, in principle, is also measurable in experiments that record real-space images of 
bosons using a quantum microscope~\cite{BGP2009,BP2010,SW2010, GB2021, BHM2024}. For example, one just needs to 
have many samples of the Bose gas in real space. From each sample, one can compute $I$ using Eq.~\eqref{eq:MI} and then 
obtain a histogram of $I$. Our theoretical predictions for the distribution of $I$ for noninteracting bosons in 
a trap can then be used as a benchmark to compare such possible experimental data, since in real systems the 
bosons are typically interacting.


Let us briefly summarise our main results. We consider $N$ noninteracting bosons in a harmonic trap $V(r) = 
(1/2) m \omega^2 {\vec r}^{\,2}$ in $d$ dimensions. The appropriate thermodynamic limit corresponds to taking 
$N \to \infty$ and $\omega \to 0$ keeping $\rho = N\,\omega^d$ fixed. In this thermodynamic limit, the system 
undergoes the BEC transition for $d>d_c=1$ across the critical density
\bea \label{rhoc_d_intro}
\rho_c(\beta) = \frac{\zeta(d)}{\beta^d} \;,
\eea
where $\zeta(n) = \sum_{k\geq 1} 1/k^n$ is the Riemann zeta function (with $n>1$) and $\beta$ denotes the inverse temperature. 
For $\rho<\rho_c(\beta)$, the system is in the fluid phase, while for $\rho > \rho_c(\beta)$ it is a condensate. 
\mk{Even though the condensation occurs in all $d>1$, we show  that the nature of the condensation
is different for $1<d\leq 2$ and $d>2$. For simplicity, the analysis for $\rho$ close to $\rho_c(\beta)$ is presented only for $d>2$.} 

We compute exactly the distribution $P_\beta(I,N)$ of the moment of inertia $I$ in the thermodynamic limit and 
show that it admits a large deviation form 
\bea \label{LDF_intro}
P_\beta(I,N) \approx e^{-V\, \Phi\left( \frac{I}{V}\right)} \;,
\eea
where $V = \omega^{-d}\gg 1$ is a large parameter and $\Phi(z)$ denotes the rate function 
associated to the distribution of $I$. We show that, whenever the system of bosons exhibits a BEC transition 
in the thermodynamic limit (e.g., when $d>1$), the rate function $\Phi(z)$ associated with $I$ has a 
singularity at a critical point $z_c$ where its second derivative is discontinuous. 
If there is no BEC transition (e.g., in $d = 1$), the rate function $\Phi(z)$ becomes an analytical function 
with no singularity. Thus, by measuring the distribution of $I$, one can detect if the system undergoes 
a BEC transition or not.  

The rest of the paper is organised as follows. In Section \ref{sec:model}, we introduce the model precisely and 
derive the identity that relates the cumulant generating function of $I$ with the canonical partition function 
$Z(N,\beta)$. In Section \ref{sec:Z} we derive the large $N$ asymptotic behaviour of $Z(N,\beta)$ in any 
dimension and show that, for $d>1$, it does have a BEC transition in the thermodynamic limit. These large $N$ 
asymptotics of $Z(N,\beta)$ are then used in Subsection \ref{sub:psi} to derive the cumulant generating 
function of $I$, using the identity established in Section \ref{sec:model}. In Subsection \ref{sub:phi}, we 
extract the rate function $\Phi(z)$ from the cumulant generating function and derive in particular its 
nonanalytic behavior at the critical point $z=z_c$. We conclude with a summary and outlook in Section 
\ref{sec:conclusion}. Some details of the computations are relegated to the three appendices.

\section{Model and the observable of interest}\label{sec:model}

We consider a system of $N$ noninteracting bosons in a harmonic trap in dimension $d$. The Hamiltonian of the
system is given by
\bea \label{H_N}
\hat H_N(m) = \sum_{i=1}^N \hat h_i(m) \quad {\rm where} \quad \hat h_i(m) = \frac{\vec{p}_i^{\,2}}{2m} + \frac{1}{2} m \omega^2 \vec{r}_i^{\,2} \;,
\eea
where $m$ is the mass of the particles and $\omega$ is the frequency of the trap. Here $\vec{p}_i$ and $\vec{r}_i$ denote respectively the 
position and the momentum of particle $i$ in dimension $d$. This is slightly different from the ideal Bose gas in a box of linear size $L$, 
which is a textbook example. There, in the thermodynamic limit, where $N \to \infty$, $L \to \infty$ but with the ratio $\rho = N/L^d$ 
fixed, it is well known~\cite{Ziff77,Huang_book,Pathria_book} that the system undergoes a condensation transition -- the celebrated Bose-Einstein condensation (BEC) -- when the 
density $\rho$ exceeds a critical value $\rho_c$ at a fixed temperature, or equivalently when the temperature goes below a critical 
temperature $T_c$ for a fixed density. For $\rho<\rho_c$ the system is in a ``fluid'' phase, while for $\rho > \rho_c$, it is in a 
``condensed'' phase where an extensive number of Bosons condense in the single-particle ground state. For the box geometry, this happens for 
any dimension $d >2$~\cite{Huang_book,Pathria_book}. In contrast, for a harmonically trapped Bose gas, there is no confining volume. Indeed, from Eq. (\ref{H_N}), it is 
clear that $1/\omega$, which typically denotes the width of the potential, plays an analogous role as the size $L$ in the box geometry. 
Hence, in this case $V=1/\omega^d$ is the analogue of the volume in the box geometry.
Consequently the thermodynamic limit here corresponds to taking the 
limit $N \to \infty$, $\omega \to 0$ but keeping the product
\bea \label{def_rho}
\rho  =N \omega^d \;,
\eea 
fixed (see e.g. \cite{Satya_Bose96}). Henceforth we will call this product $\rho$ as the ``density''. As in the box geometry, there is a BEC 
transition in the harmonic case also, as one increases the density $\rho$ through a critical density $\rho_c(\beta)$ for a fixed 
inverse temperature $\beta$ 
(alternatively reducing the temperature through a critical temperature $T_c$ for a fixed density). However, in contrast to the box case, the 
BEC transition in the harmonic case occurs for any dimension $d>1$~\cite{Stringari99}.

Our goal in this paper is to compute the statistics of the observable 
\bea \label{def_I}
I = m\,\omega^2 \sum_{i=1}^N \vec{r}_i^{\,2} \;,
\eea
at any temperature $T$ and density $\rho$ for a harmonically trapped ideal Bose gas. This observable represents twice the potential energy of the Bose gas defined in Eq. (\ref{H_N}). We will show that the harmonic case allows an exact solution for the distribution of $I$ in the thermodynamic limit. 
The distribution of $I$ is defined as follows. Consider $N$ noninteracting bosons at a fixed total energy $E$. The system is fully specified in real space provided we know the many-body bosonic eigenfunction $\Psi_E(\vec{r}_1, \cdots, {\vec r_N})$, normalized to unity. Here ${\vec r}_i$ represents the positon of the $i$-th boson. By definition, $|\Psi_E(\vec{r}_1, \cdots, {\vec r_N})|^2$ represents the quantum joint probability density arising purely from quantum fluctuations at a fixed energy $E$. Of course the energy itself fluctuates at a fixed temperature. 
In the canonical ensemble where the temperature $T$ and the number of bosons $N$ is fixed, both the quantum and thermal fluctuations are captured by the joint probability density function (JPDF) 
\bea \label{JPDF}
P_\beta({\vec r}_1, \cdots, {\vec r}_N) = \frac{1}{Z(N,\beta)} \sum_{E}e^{- \beta E}\, |\Psi_E(\vec{r}_1, \cdots, {\vec r_N})|^2 \;.
\eea 
Here the partition function $Z(N,\beta)$ ensures the normalization of the JPDF and $\beta = 1/T$ represents the inverse temperature. The total energy $E$ can be conveniently expressed in terms of the occupation numbers $n_{\vec j} = 0,1,2, \cdots$ of the single-particle energy levels $\epsilon_{\vec j}$ labeled by the $d$-dimensional vector ${\vec j} = (j_1, \cdots, j_d)$ and given by
\bea \label{energy}
\epsilon_{\vec j} = \left(j_1 + j_2+ \cdots + j_d + \frac{d}{2}\right) \hbar \omega \;,
\eea
where $j_i$'s are nonnegative integers. The total energy $E$ of a many-body state and the total number of particles can be expressed in terms of the occupation numbers $n_{\vec j}$'s
\bea \label{eq} \label{def_E}
E = \sum_{\vec j}\,  n_{\vec j} \, \epsilon_{\vec j} \quad, \quad {\rm} \quad \quad N = \sum_{\vec j}\, n_{\vec j}  \;.
\eea
The canonical partition function $Z(N,\beta)$ is then given by
\bea \label{def_Z}
Z(N,\beta) = \sum_{E} e^{- \beta\, E} = \sum_{\{n_{\vec j}\}} e^{- \beta \sum_{\vec j} n_{\vec j} \epsilon_{\vec j}}  \, \delta_{\sum_{\vec j} n_{\vec j},N} \;.
\eea
Taking the generating function with respect to (w.r.t.) $N$ one finds
\bea \label{GF_Z}
\sum_{\mk{N\geq 0}} z^N Z(N,\beta) = \prod_{{\vec j}} \frac{1}{1-z\,e^{- \beta \epsilon_{\vec j}}} \,,
\eea
where the product runs over all possible vectors ${\vec j} = (j_1, \cdots, j_d)$ where $j_i$'s are nonnegative integers. Clearly the ground state corresponds to ${\vec j}= {\vec 0} = (0,0, \cdots, 0)$. Inverting Eq. (\ref{GF_Z}) formally, using Cauchy's formula, gives
\bea \label{invert_Z}
Z(N,\beta) = \oint \frac{dz}{2\pi i}  \frac{1}{z^{N+1}}\prod_{{\vec j} } \frac{1}{1-z\,e^{- \beta \epsilon_{\vec j}}} \;,
\eea
where the contour integral over $z$ runs over a circle centered 	at $z=0$. Making further the change of variable $z=e^{-s}$ one can convert it to a Bromwich integral in the complex $s$-plane, namely
\bea \label{Bromw}
Z(N,\beta) = \int_{\Gamma} \frac{ds}{2\pi i}\, e^{s N} \frac{1}{1-e^{-s - \beta \epsilon_{\vec 0}}} e^{-\sum_{\vec j \neq \vec 0} \ln{(1 - e^{-s -\beta \epsilon_{\vec j}})}} \;,
\eea
where we have separated out the contribution from the ground state with energy 
\bea \label{GS}
\epsilon_{\vec 0} = \hbar \omega \frac{d}{2} \;,
\eea
corresponding to $j_1 = j_2 = \cdots = j_d = 0$. In Eq. (\ref{Bromw}), the Bromwich contour $\Gamma$ runs vertically along the imaginary axis whose real part lies to the right of all the singularities of the integrand. 

Given the joint distribution at finite inverse temperature $\beta$ given in Eq. (\ref{JPDF}), the probability distribution of the linear statistics $I$ defined in Eq. (\ref{def_I}) can be expressed as
\bea \label{PDF_I}
P_\beta(I,N) = \int d{\vec r}_1 \cdots \int d{\vec r}_N \, P_\beta({\vec r}_1, \cdots, {\vec r}_N) \, \delta\left(I - m\,\omega^2  \sum_{i=1}^N \vec{r}_i^{\,2}   \right) \;.
\eea
Taking the Laplace transform w.r.t. $I$ gives
\bea \label{Laplace_PDF_I}
\tilde P_\beta(\lambda,N) &=& \int_0^\infty \, e^{-\lambda\, I} \, P_\beta(I,N) \, dI = \int d{\vec r}_1 \cdots \int d{\vec r}_N \, P_\beta({\vec r}_1, \cdots, {\vec r}_N)\, e^{-\lambda m\,\omega^2 \sum_{i=1}^N  \vec{r}_i^{\,2}} \; \nonumber \\
&=& \frac{1}{Z(N,\beta)} \sum_{E}\, e^{- \beta E}  \int d{\vec r}_1 \cdots \int d{\vec r}_N\, |\Psi_E(\vec{r}_1, \cdots, \vec{r}_N)|^2\, e^{-\lambda m\,\omega^2 \sum_{i=1}^N  \vec{r}_i^{\,2}} \;,
\eea
where we used the explicit expression of $P_\beta(\vec{r_1}, \cdots, \vec{r}_N)$ in Eq. (\ref{JPDF}). In general, this multi-dimensional 
integral is hard to perform explicitly, since the many-body wave function of an excited state $E$ is a permanent of the associated 
single-particle eigenfunctions (each of them being the product of a Gaussian and a Hermite polynomial, one  for each coordinate). Moreover, 
summing over all possible energies $E$ makes the evaluation of $\tilde P_\beta(\lambda,N)$ in Eq. (\ref{Laplace_PDF_I}) rather daunting. To 
circumvent this problem, we make use of a nontrivial identity, which was first derived for one-dimensional noninteracting fermions in 
\cite{Grela_PRL} by a rather complicated method. Here, we first rederive this identity by a much simpler method
that demonstrates its generality, namely that it is valid in all dimensions and at all 
temperatures, irrespective of the statistics of the particles, i.e., it holds for both fermions and bosons. In Appendix \ref{app_duality} we 
provide a simple proof of this identity which reads
\bea \label{relation_tilde}
\tilde P_\beta(\lambda,N) = \langle e^{-\lambda m\, \omega^2  \sum_{i=1}^N \vec{r}_i^{\,2}}\rangle = \frac{Z(N,\tilde \beta(\lambda))}{Z(N,\beta)} \;,
\eea 
where $Z(N,\beta)$ is the canonical partition function given in Eq. (\ref{Bromw}) and $\tilde \beta(\lambda)$ is determined
from the exact relation
\bea  \label{rel_btilde}
 \cosh{(\tilde \beta(\lambda) \hbar \omega)} = \cosh{(\beta \hbar \omega)}  + \lambda \,{\hbar \omega} \,  \sinh{(\beta \hbar \omega)} \;.
\eea
In Eq. (\ref{relation_tilde}), $\langle \cdots \rangle$ denotes an average over the JPDF in Eq. (\ref{JPDF}). Note that when $\lambda=0$, we have $\tilde \beta(0)=\beta$.
This identity relates the statistics of $I$ to the partition function $Z(N,\beta)$ in Eq. (\ref{Bromw}), the computation of which does not 
require any knowledge of the many-body eigenfunction $\Psi_E(\vec{r}_1, \cdots, \vec{r}_N)$. Hence, thanks to this identity 
(\ref{relation_tilde}), the statistics of the operator $I$ depend only on the spectrum of $\hat H_N$ (i.e., its energy eigenvalues) but not 
on its eigenfunctions. We call this identity in Eqs. (\ref{relation_tilde}) and (\ref{rel_btilde}) a duality relation as
it involves partition functions at two different temperatures. {Let us remark that the emergence of the inverse effective temperature $\tilde\beta(\lambda)$ is a highly non-trivial consequence of the precise form of the operator representing the moment of inertia and also the quantum propagator of the harmonic oscillator (see Appendix.~\ref{app_duality}). 
This mechanism behind this re-normalized temperature is reminiscent of the `tilted' effective potential that appears in the study of linear statistics in Coulomb gases in a confining potential \cite{Tilted_CG24}. The cumulant generating function of such linear statistics in confined Coulomb gases gives rise to an additional term in the action that renormalizes, i.e., tilts the confining potential by an additional $\lambda$-dependent tilted potential that depends on the linear statistics.}

Given this relation (\ref{relation_tilde}), one can compute any moments of $I$ by taking repeated derivatives of (\ref{relation_tilde}) w.r.t. $\lambda$ and setting $\lambda= 0$. For example, the average $\langle I \rangle$ is simply given by
\bea \label{av_I} 
\langle I \rangle = - \frac{\partial}{\partial \lambda} \langle e^{-\lambda I }\rangle\Big \vert_{\lambda=0} \;.
\eea
Taking the derivative of Eq. (\ref{relation_tilde}) w.r.t. $\lambda$ and using the relation (\ref{rel_btilde}) and setting $\lambda = 0$ gives the exact relation
\bea \label{av_I_2}
\langle I \rangle = - \frac{\partial \ln Z(N,\beta)}{\partial \beta} \;.
\eea
The right hand side represents just the average energy of the system (up to a factor $1/m$). This is expected since, for the harmonic oscillator, the kinetic part and the potential part of the total energy have identical behaviors (since the position ${\vec r}_i$ and the momentum ${\vec p}_i$ play a symmetric role in the Hamiltonian). Hence the average total energy is twice that of the potential energy and hence coincides with the average value of the observable $I = m\,\omega^2 \sum_{i=1}^N \vec{r}_i^{\,2}$. 

Similarly, by taking two derivatives of Eq. (\ref{relation_tilde}) w.r.t. $\lambda$, one can relate the variance of $I$ to the partition function as follows (for a detailed derivation, see Appendix \ref{App_var})
\bea \label{rel_var}
{\rm Var}(I) = \langle I^2 \rangle -  \langle I \rangle^2 = \left[ \frac{\partial^2 \ln Z}{\partial \beta^2} + {\hbar \omega}\, {\rm coth}(\beta \hbar \omega) \langle I \rangle\right] \;.
\eea
Note that the first term on the right hand side is exactly the specific heat of the system, while the second term represents a quantum correction.

We thus see from Eq. (\ref{relation_tilde}) that computing the full distribution $P_\beta(I,N)$ requires the knowledge of the exact partition function $Z(N,\beta)$ as a function of $\beta$ for a given $N$. Analysing the partition function $Z(N,\beta)$ for finite $N$ turns out to be rather hard. However, in the next section, we show that we can make progress in the appropriate large $N$ limit mentioned in Eq. (\ref{def_rho}). Namely we will take the limits 
\bea \label{limits}
N \to \infty \quad  {\rm and} \quad  V = \frac{1}{\omega^d} \to \infty\;, \quad {\rm while \; keeping \; the \;  ratio} \; \rho = \frac{N}{V} = N \omega^d \; \; {\rm fixed} \;.
\eea

\section{Large $N$ analysis of the partition function $Z(N,\beta)$}\label{sec:Z}

For simplicity, we will henceforth use units in which $m = 1$ and $\hbar = 1$. Our starting point is the exact expression for 
$Z(N,\beta)$ in Eq. (\ref{Bromw}) which is valid for arbitrary $N$. We first rewrite this expression as
\bea \label{ZN_sp}
Z(N,\beta) = \int_{\Gamma} \frac{ds}{2\pi i}\, \frac{1}{1-e^{-s - \beta \epsilon_{\vec 0}}} e^{N\, s - g_N(\beta,s)} \quad, \quad 
g_N(\beta,s) = \sum_{\vec j \neq \vec 0} \ln{(1 - e^{-s -\beta \epsilon_{\vec j}})} \;.
\eea 
In the scaling limit $\omega \to 0$, one can replace the sum in $g_N(\beta,s)$ by an integral. 
This is done as follows. We first define the variables 
\bea \label{def_ki}
k_i = \beta \omega j_i \;,
\eea
where $j_i$'s are non-negative integers. In the limit $\omega \to 0$, the $k_i$'s become essentially continuous variables with increment 
$\Delta k_i = \beta \omega \to 0$. Therefore, multiplying and dividing $g_N(\beta,s)$ in Eq. (\ref{ZN_sp}) by $(\beta \omega)^d$, we can 
replace the sum by a $d$-dimensional integral (valid in the limit $\omega \to 0$)
\bea \label{gN_1}
g_N(\beta,s) \approx \frac{1}{(\beta \omega)^d} \int_> d \vec{k} \, \ln\left(1 - e^{-s - \sum_{i=1}^d k_i} \right)  = \frac{V}{\beta^d}  \int_> d \vec{k} \, \ln\left(1 - e^{-s - \sum_{i=1}^d k_i} \right) \;,
\eea
where $V = 1/\omega^d$ is a fixed number of order $O(1)$, as mentioned in Eq. (\ref{def_rho}). The 
integral in Eq. (\ref{gN_1}) runs over the ``first quadrant'' in the $d$-dimensional Fourier space, i.e., 
over $k_i \geq 0$, as indicated by the notation $\int_>{d \vec k}$. Furthermore, in the limit 
$\beta \omega \to 0$, the term in the integrand in Eq. (\ref{ZN_sp}) representing the contribution from the ground state reduces 
simply to ${1}/{(1-e^{-s - \beta \epsilon_{\vec 0}})} \approx 1/(1-e^{-s})$. Therefore in this limit, using $N=\rho\, V$, the partition function reduces to
\bea \label{ZN_sp2}
Z(N,\beta) = \int_{\Gamma} \frac{ds}{2\pi i}\, \frac{1}{1-e^{-s}} \, e^{V A(s,\rho)} \quad, \quad 
A(s,\rho) = s \rho - \frac{1}{\beta^d} \int_> d\vec k \ln \left(1 - e^{-s - \sum_{i=1}^d k_i} \right) \; .
\eea

\begin{figure}[t]
\centering
\includegraphics[width = 0.5\linewidth]{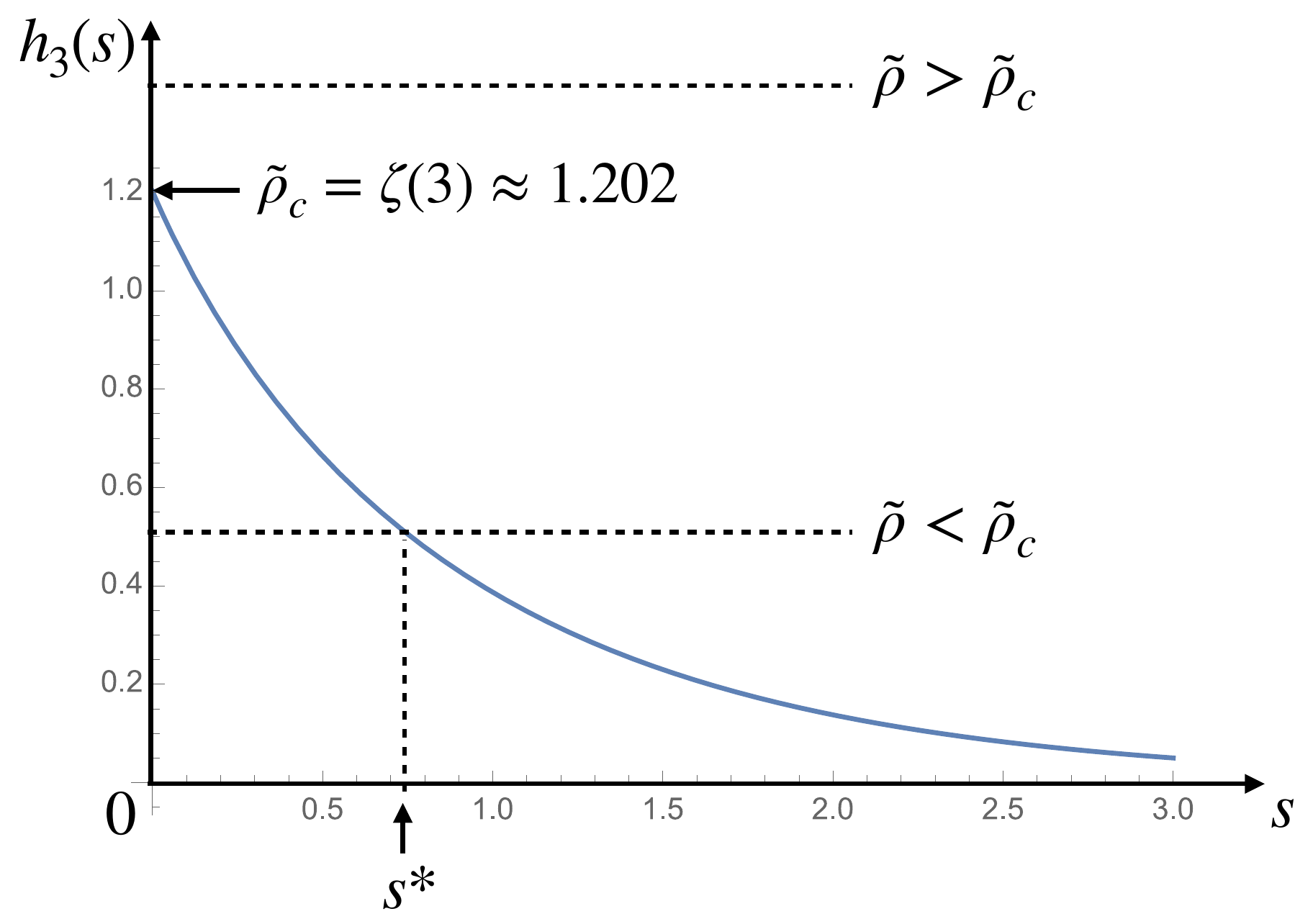}
\caption{The solid blue line is a plot of the 
function $h_{d=3}(s) = {\rm Li}_{d=3}(e^{-s})$ given in Eq. (\ref{sp_eq2}) as a function of $s$. The value at $s=0$, namely $h_{d=3}(s=0) = \zeta(3)$, denotes the critical density $\tilde \rho_c$. The two horizontal lines {indicate} two representative values of the density $\tilde \rho$: one in the fluid phase ($\tilde \rho < \tilde \rho_c$) and the other in the condensed phase (where $\tilde \rho > \tilde \rho_c$). In the fluid phase, given the density $\tilde \rho < \tilde \rho_c$, there is a unique value $s^*$ such that $h_{d=3}(s^*) = \tilde \rho$ as shown in the figure. This shows that, in the fluid phase, there is a nonzero value $s=s^*$ where a saddle point occurs. As $\tilde \rho$ approaches $\tilde \rho_c$ from below, the saddle point $s^*$ approaches to~$0$.}\label{Fig_rho}
\end{figure}

\begin{figure}[t]
\centering
\includegraphics[width = 0.45\linewidth]{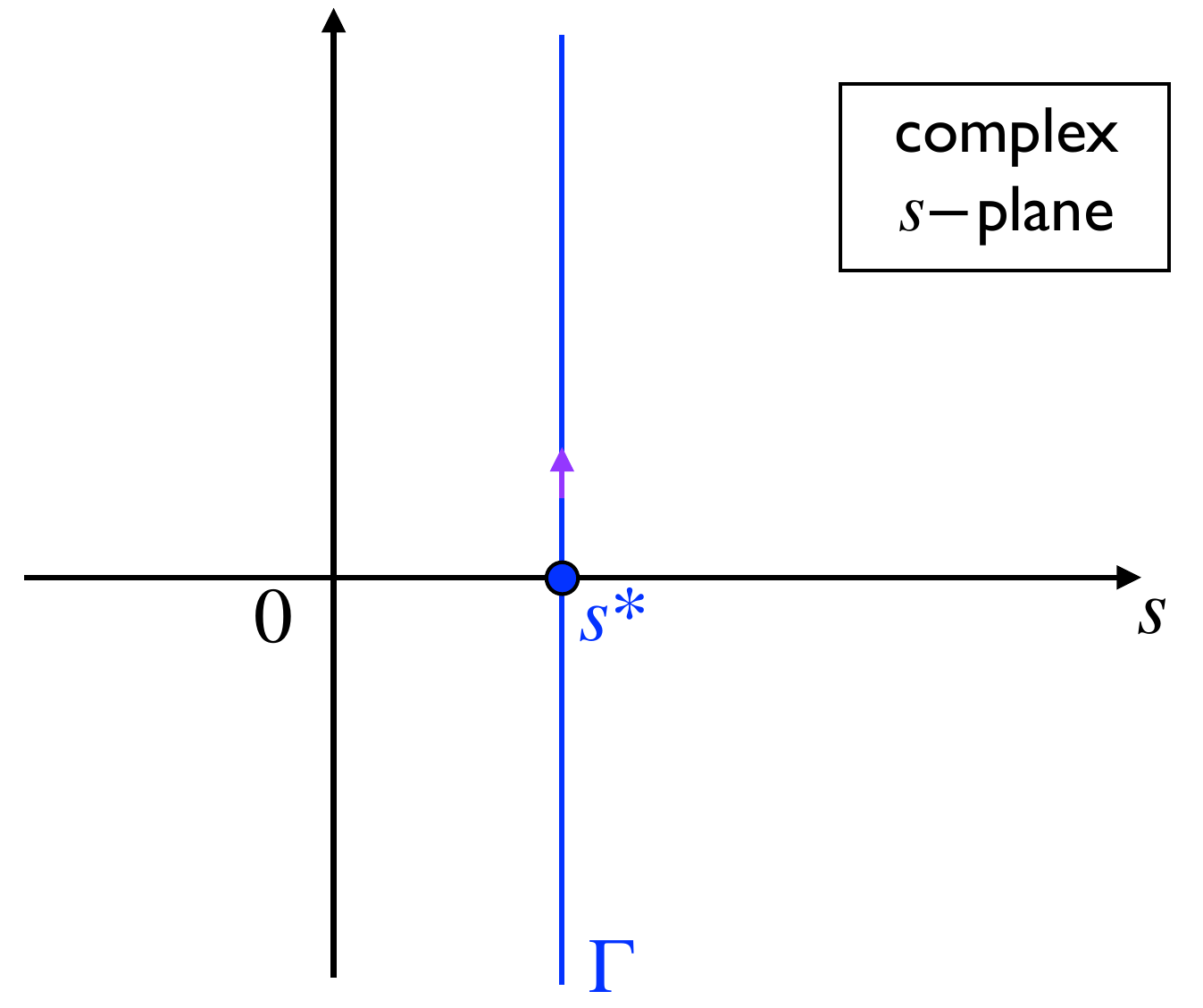}
\caption{The vertical blue line denotes the Bromwich contour $\Gamma$ in 
the complex $s$-plane in Eq. (\ref{ZN_sp2}). The Bromwich contour intersects the real axis at $s=s^*$ denoting the location of the saddle point in Eq. (\ref{sp_eq2}).}\label{Fig_Brom}
\end{figure}

Before proceeding, let us make an important remark: the integrand in the expression of $Z(N,\beta)$ in
Eq. (\ref{ZN_sp2}) has a pole at $s=0$ arising from the ground state contribution. Hence the real part 
of the Browmich contour must be at $s>0$. The structure of the integral clearly suggests to look for a saddle point solution on the real 
axis at $s=s^*>0$. Setting $\partial A/\partial s = 0$ at $s=s^*$ one gets the saddle point equation
\bea \label{sp_eq}
\rho = \frac{1}{\beta^d}\int_> d\vec k \frac{1}{e^{s^*+\sum_{i=1}^d k_i}-1} \;.
\eea
The integral on the right hand side can be performed explicitly in $d$ dimensions and it leads to (see Appendix \ref{App_kintegral})
\begin{equation} \label{sp_eq2}
\tilde \rho = \rho \, \beta^d = \frac{1}{\Gamma(d)} \, \int_0^{\infty} dq\, q^{d-1}  \frac{1}{e^{s^*+q}-1}\, =  
h_d(s^*) \equiv {\rm Li}_{d}(e^{-s^*})\, , \,\, {\rm where} \quad {\rm Li}_n(z) = \sum_{k=1}^\infty \frac{z^k}{k^n}\;.
\end{equation}
The function $h_d(s) \equiv {\rm Li}_d(e^{-s})$ is a monotonically decreasing function of $s$ 
(see Fig. \ref{Fig_rho} for $d=3$) with asymptotic behaviors given by
\bea \label{h_asympt}
h_d(s) \approx
\begin{cases}
& \zeta(d) \quad, \quad s \to 0 \\
& \\
& e^{-s} \quad, \quad s \to \infty \;.
\end{cases}
\eea
Thus the saddle point $s^*$ is given by the positive root of $h_d(s^*) = \tilde \rho$. Since $h_d(s)$ is a monotonically decreasing function 
of $s$, its maximum value for nonnegative $s$ occurs at $s=0$, with the value $h_d(0) = \zeta(d)$. We note that $\zeta(d)$ is finite only 
for $d>1$. For $d \leq 1$, $h_d(s) \to + \infty$ as $s \to 0$. Thus for $d \leq 1$, there will always be a saddle point solution $s^* >0$. 
In contrast, for $d>1$, a saddle point solution may or may not exist, depending on the value of $\tilde \rho = \rho\beta^d$, which is our 
control parameter. For $d>1$, since $h_d(0) = \zeta(d)$ is finite, there will be a saddle point solution $s^*>0$ provided $\tilde 
\rho < \tilde \rho_c$ (see Fig. \ref{Fig_rho} for an illustration in $d=3$) where
\bea \label{rhoc_d}
\tilde \rho_c =  \zeta(d) \;.
\eea
This defines a critical line $\rho\, \beta^d = \zeta(d)$ in the $(\rho,\beta)$ plane across which the BEC occurs. This critical line can be 
traversed either by changing $\rho$ at fixed $\beta$ or by changing $\beta$ at fixed $\rho$. Below, for convenience, we will vary $\rho$ 
through this critical line, with $\beta$ fixed. In this case the critical density $\rho_c(\beta)$ is given by
\bea \label{rhoc}
\rho_c(\beta) = \frac{\zeta(d)}{\beta^d} \;.
\eea
As $\rho$ approaches $\rho_c(\beta)$ from below, the saddle point $s^* \to 0$ (see Fig. \ref{Fig_Brom}).  For $\rho > \rho_c(\beta)$, there 
is thus no saddle point and one has to evaluate the Bromwich integral along the imaginary axis passing through $s \to 0^+$. Thus for $d>1$, 
as one varies $\rho$ through the critical value $\rho_c(\beta)$ the system undergoes a transition. We 
will see later that the phase $\rho < \rho_c(\beta)$ where there is a positive saddle point $s^*>0$, corresponds to the so called ``fluid'' 
phase. In contrast, for $\rho > \rho_c(\beta)$ when there is no saddle point, the system is in a Bose-Einstein condensed (BEC) phase.

For later purposes we note that the action $A(s,\rho)$ in Eq. (\ref{ZN_sp2}) can also be evaluated explicitly by performing the integral over $\vec{k}$ (following the steps in  Appendix \ref{App_kintegral}), leading to 
\bea \label{A_expl}
A(s,\rho) = \rho s + \frac{1}{\beta^d}{\rm Li}_{d+1}(e^{-s}) \;.
\eea 
We will focus here only on $d>1$ when there is a condensation transition, as argued above, at a critical density $\rho = \rho_c(\beta) = \zeta(d)/\beta^d$. 
For later purposes, it will also be useful to extract the small $s$ expansion of $A(s,\rho)$ in Eq. (\ref{A_expl}).

For noninteger $d>1$, it reads
\bea  \label{A_small_s_nonintd}
A(s,\rho) = \left[\frac{1}{\beta^d}\zeta(d+1) + \left(\rho - \rho_c(\beta) \right)\,s + \frac{1}{\beta^d}\zeta(d-1)\, s^2  + \cdots\right] + \frac{\Gamma(-d)}{\beta^d}\,s^d + \cdots \;.
\eea
where the square brackets indicate the regular part of the expansion, while $\Gamma(-d)\,s^d$ is the leading singular term in the small $s$ expansion. For integer $d>1$, \mk{we similarly have
a regular and a singular part that read}
\begin{equation} \label{A_small_s_intd}
A(s,\rho) = \left[\frac{1}{\beta^d}\zeta(d+1) + \left(\rho - \rho_c(\beta) \right)\,s + \mk{O(s^2)}  \right] +  \frac{(-1)^{d+1}}{\beta^d\,\Gamma(d+1)}\, s^{d}\,\ln s + \cdots \;,
\end{equation}
where $\rho_c(\beta)= \zeta(d)/\beta^d$ for all $d>1$ denotes the critical density. 

\mk{It turns out that the leading small $s$ behavior of $A(s,\rho)$ that contributes dominantly to the saddle point depends crucially on the dimension $d$. There are three different cases which we consider here separately:
\begin{enumerate}
\item{The case $1<d<2$: in this case, the three leading terms that contribute dominantly to the saddle point are
given by
\bea \label{A1d2}
A(s,\rho) \approx \frac{\zeta(d+1)}{\beta^d}\, + (\rho-\rho_c(\beta))\, s
+ \frac{\Gamma(2-d)}{(d-1)\,d\, \beta^d}\, s^d \;,
\eea
where we used the relation $\Gamma(1+x) = x \Gamma(x)$ to rewrite the singular term in (\ref{A_small_s_nonintd}).
}
\item{The case $d=2$: in this case, the three leading terms are given by
\bea \label{Asd2}
A(s,\rho) \approx \frac{\zeta(3)}{\beta^2}  + (\rho-\rho_c(\beta))\, s -\frac{1}{2\beta^2}\, s^2\, \ln(s) \;.
\eea
}
\item{The case $d>2$: in this case, we need to use the expansion for small $s$ as shown in Eq. (\ref{A_small_s_nonintd}) for non-integer $d$ and Eq. (\ref{A_small_s_intd}) for integer $d$. 
} 
\end{enumerate}
}

\mk{Thus the nature of the first nonlinear term in the small $s$ expansion is different in
the three cases. It is singular for $1<d\le 2$ and regular (quadratic) for $d>2$. For $1<d<2$, it
has a power law singularity, while exactly at $d=2$ it has a logarithmic singularity.
It turns out that the nature of the condensation transition at $\rho=\rho_c(\beta)$ is controlled
by this leading nonlinear term. Hence the condensation transition is
quite different for $1<d<2$ where it is `anomalous' due to the singular term
in the small $s$ expansion in Eq. (\ref{A1d2}). In contrast, for $d>2$, it is `normal'
due to the regular quadratic form of the first nonlinear term. The case $d=2$ is
at the borderline. A similar `anomalous'
vs. `normal' condensation, as one changes a parameter (in our case the dimension $d$), 
was also seen in various models of real space condensation in
mass transport models such as zero range processes~\cite{MEZ05,EMZ06}, as well as in models of run-and-tumble particles~\cite{gradenigo2019,mori2021,mori2021b}.
We borrow this terminology `anomalous' vs. `normal' from this literature.
The analysis of the partition function
$Z(N,\beta)$ and other observables in the harmonically trapped Bose gas 
near the transition at $\rho=\rho_c(\beta)$ 
turns out to be a bit more involved in the
anomalous case namely for $1<d\le 2$, compared to the normal case for $d>2$. Hence,
to keep the analysis of the nature of the transition near $\rho=\rho_c(\beta)$ simple, 
we will restrict ourselves in this paper to $d>2$ where the condensation is `normal'.
However, our results can be generalised easily to the anomalous case where $1<d\le 2$ \cite{HKMS}.}

\subsection{The fluid phase $\rho < \rho_c(\beta)$: when there is a saddle point}

In this case, for any $d>1$, there is a saddle point which is obtained by minimizing the action $A(s,\rho)$ in Eq.~(\ref{ZN_sp2}) as a function of $s$ for a fixed $\rho < \rho_c(\beta)$ (and fixed $\beta$).  
This leads to the explicit saddle point equation (\ref{sp_eq2}) which can then be used to obtain $s^*(\rho)$ by inverting the relation 
\bea  \label{eq_sp_A2}
{\rm Li}_d(e^{-s^*}) = \rho\beta^d \;. 
\eea
Substituting this value of $s^*$ in the action $A(s,\rho)$ in Eq. (\ref{A_expl}) gives 
$A(s^*(\rho),\rho)$. Consequently, evaluating the partition function in Eq. (\ref{ZN_sp2}) for large $N$, we get
\bea \label{ZN_sp3}
\frac{\ln Z(N,\beta)}{V} \underset{V \to \infty}{\longrightarrow} A(s^*(\rho),\rho) =  \rho s^* + \frac{1}{\beta^d}{\rm Li}_{d+1}(e^{-s^*})\;,
\eea
where we have used the explicit expression of $A(s,\rho)$ from Eq. (\ref{A_expl}). Even though this result is valid for all $\rho < \rho_c(\beta)$ in the fluid phase, one can obtain a more explicit expression for $Z(N,\beta)$ in the vicinity of the critical point, i.e., when $\rho$ approaches $\rho_c(\beta)$ from below. As $\rho \to \rho_c(\beta)$ the saddle point $s^*(\rho) \to 0$ in Fig. \ref{Fig_Brom}. 
In this limit, \mk{and restricting ourselves to $d>2$ for simplicity}, it is easy to show that the action $A(s^*,\rho)$ in (\ref{ZN_sp3}) behaves quadratically as a function of $(\rho -\rho_c(\beta))$ for small $(\rho -\rho_c(\beta))$ (see Fig. \ref{Fig_logZ}), namely
\bea \label{quadratic}
\frac{\ln Z(N,\beta)}{V} \underset{V \to \infty}{\longrightarrow} 
A(s^*,\rho) \approx \frac{\zeta(d+1)}{\beta^d}  - \frac{\beta^d}{\mk{2} \zeta(d-1)} (\rho_c(\beta)-\rho)^2  \;.
\eea
The same result can also be obtained more easily by first expanding $A(s,\rho)$ for small $s$ as in Eqs. (\ref{A_small_s_nonintd}) and (\ref{A_small_s_intd}), respectively for noninteger $d>2$ and integer $d>2$, and then minimizing with respect to $s$.

\subsection{The condensed phase $\rho > \rho_c(\beta)$: when there is no saddle point}

We start our analysis here with $d>2$ for simplicity. When $\rho$ approaches $\rho_c(\beta)$ from below, the saddle point $s^* \to 0$. We recall that, since the integrand in Eq. (\ref{ZN_sp2}) has a pole as $s=0$, the real part of the Bromwich contour can not cross $s=0$. Hence one has to analyse the integral in Eq. (\ref{ZN_sp2}) along a contour that passes just to the right of $s=0$. In other words, we need to use the small $s$ expansion of $A(s,\rho)$ in Eqs. (\ref{A_small_s_nonintd}) and (\ref{A_small_s_intd}) and then analyse the integral, but without the existence of a saddle point. In fact, from the small $s$ expansion in Eqs. (\ref{A_small_s_nonintd}) and (\ref{A_small_s_intd}), it is clear that the function $A(s,\rho)$ has a minimum at $s=s^*$ only for $\rho<\rho_c(\beta)$. For $\rho > \rho_c(\beta)$ and using the $s \to 0$ expansion of $A(s,\rho)$ in Eqs. (\ref{A_small_s_nonintd}) and (\ref{A_small_s_intd}), we first rewrite the expression of the partition function as
\bea
 \label{ZN_sp5}
Z(N,\beta) \approx e^{V \frac{\zeta(d+1)}{\beta^d}}\, \int_{\Gamma} \frac{ds}{2\pi i}\, \frac{1}{s} \, e^{V\left( \rho - \rho_c(\beta)\right)s + \frac{\zeta(d-1)}{\mk{2}\beta^d}V\,s^2 + \cdots} \;,
\eea 
\mk{where the `$\cdots$' denote higher order terms that become negligible in the large $V$ limit.} Consider now $(\rho - \rho_c(\beta)) \sim O(1)$, while $V \to \infty$. In this scale, by rescaling $s\,V = \tilde s$, we see that only the leading term inside the exponential survives (all higher order terms, including the quadratic one vanishes in this limit). This leads to
 \bea
 \label{ZN_sp5_2}
Z(N,\beta) \approx e^{V \frac{\zeta(d+1)}{\beta^d}}\, \int_{\Gamma} \frac{d\tilde s}{2\pi i}\, \frac{1}{\tilde s} \, e^{\left( \rho - \rho_c(\beta)\right)\tilde s} = e^{V \frac{\zeta(d+1)}{\beta^d}}\, \theta(\rho - \rho_c(\beta))\;,
\eea 
where we used the fact that the integral over $\tilde s$ is simply $\theta(\rho-\rho_c(\beta))$, with $\theta(x)$ denoting the Heaviside step function. 
This relation (\ref{ZN_sp5_2}) can be proved by first taking a derivative with respect to $\rho - \rho_c(\beta)$, which gives $\delta(\rho-\rho_c(\beta))$ and then integrating it back. This gives, in the condensed phase where $(\rho-\rho_c(\beta)) >0$ and is $O(1)$
\bea \label{ZN_sp3_2}
\frac{\ln Z(N,\beta)}{V} \underset{V \to \infty}{\longrightarrow} \frac{\zeta(d+1)}{\beta^d} \;.
\eea
Thus this leading order behavior is independent of $\rho-\rho_c(\beta)$. {Eqs. \eqref{ZN_sp3_2} and \eqref{quadratic} show that $\ln Z(N,\beta)/V$, to leading order for large $V$, increases monotonically with $\rho$ till $\rho_c(\beta)$ (fluid phase) and then ``freezes'' to the value $\zeta(d+1)/\beta^d$ for $\rho > \rho_c(\beta)$ in the condensed phase (see Fig. \ref{Fig_logZ}). }

Identifying $-\ln Z(N,\beta)/V$ for large $V$ as the free energy per unit volume in the thermodynamic limit, denoted by
\bea \label{def_free_e}
f(\rho,\beta) = -\lim_{V \to \infty}  \frac{\ln Z(N,\beta)}{V} \;,
\eea
we now summarise how $f(\rho,\beta)$ behaves across the critical density $\rho_c(\beta) = \zeta(d)/\beta^d$. We find 
\bea \label{summary_free}
-f(\rho,\beta) = \lim_{V \to \infty} \frac{\ln Z(N,\beta)}{V} = 
\begin{cases}
& \rho s^* + \frac{1}{\beta^d}\,{\rm Li}_{d+1}(e^{-s^*})  \quad, \quad \rho < \rho_c(\beta) \; ({\rm fluid}) \\
& \\
& \frac{\zeta(d+1)}{\beta^d} \quad, \quad \hspace*{2.2cm} \rho > \rho_c(\beta) \; ({\rm condensed}) \;,
\end{cases}
\eea
where $s^*$, in the first line, is given by the solution of ${\rm Li}_d(e^{-s^*}) = \rho\,\beta^d$. Thus the free energy per unit volume in the thermodynamic limit undergoes a {\it freezing} transition as the density $\rho$ crosses from the fluid side to the condensed side. 

\mk{Let us remark that, here, 
we derived the freezing transition at $\rho = \rho_c(\beta)$ in the condensed side by analyzing the Bromwich integral at $s=0$ and restricting ourselves to $d>2$ for simplicity. However, exactly the same analysis can be carried out near $s=0$ using the appropriate small $s$ expansion of $A(s,\rho)$ for $1<d<2$. In fact, the result in Eq. (\ref{summary_free}) is valid for any $d>1$.}

Let us end this section with the following side remark. Whenever there is a saddle point in the integral in Eq. (\ref{ZN_sp2})
at a finite value $s=s^*$, i.e., the system is in the fluid phase ($\rho< \rho_c(\beta)$), the canonical and the grand-canonical ensembles are equivalent. However, in the condensed phase for $\rho>\rho_c(\beta)$, there is no longer a saddle in $A(s,\rho)$ in Eq. (\ref{ZN_sp2}) and, in that case, we have analysed the large $N$ behavior of the canonical partition function by directly evaluating the Bromwich integral near $s=0$. This is the correct procedure to follow to compute not just the leading order but also 
the subleading orders to the large $V$ behavior of the partition function $Z(N,\beta)$. This is different from the standard procedure used in textbooks \cite{Huang_book,Pathria_book} where one assumes that there is an ``effective'' grand canonical description even in the condensed side, with a new saddle point that occurs at $s^* \sim O(1/(V(\rho-\rho_c(\beta))))$. For instance, for the Bose gas in the harmonic trap this route was also recently used in Ref. \cite{Crisanti24}. There, the partition function in Eq. (\ref{ZN_sp2}) was rewritten~as
\bea \label{ZN_sp3.1}
Z(N,\beta) = \int_{\Gamma} \frac{ds}{2\pi i}\,  \, e^{V \left(A(s,\rho)- \frac{1}{V}\ln(1-e^{-s})\right)} \quad, \quad \;
\eea
and a saddle point was sought for the 
new effective action modified by the presence of the 
order $O(1/V)$ correction term. Evaluating the action at the new saddle point provides correctly the leading order of the partition function. However, the subleading corrections are not correctly captured by this effective saddle point method. For that, one needs to perform carefully the Bromwich integral along the 
vertical axis close to $s=0$, as is done here. A more concrete example of this is provided in the next subsection.

\subsection{Connecting the fluid and the condensed phase on a smaller scale of $(\rho-\rho_c(\beta)) \sim O(1/\sqrt{V})$}
\label{subsec:match}

\begin{figure}[t]
\centering
\includegraphics[width = 0.5 \linewidth]{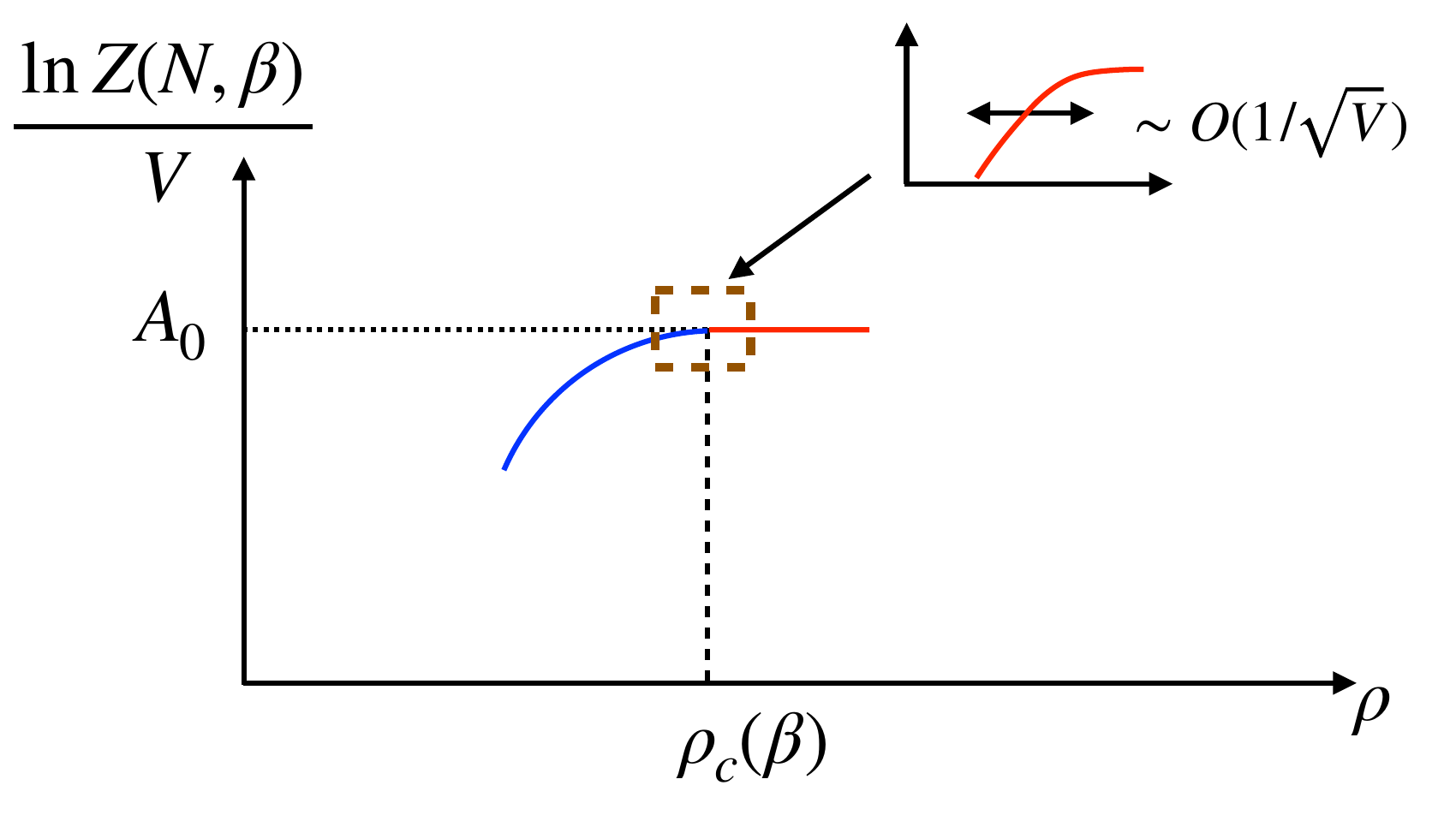}
\caption{Plot of the $\ln Z(N,\beta)/V$ (in the limit $V \to \infty$) vs $\rho$ close to the critical density $\rho = \rho_c(\beta)$ as given in Eq. (\ref{summary}). The curve decreases quadratically for $\rho<\rho_c(\beta)$ (shown by the blue line), and freezes to the constant $A_0 = \zeta(d+1)/\beta^d$ for $\rho>\rho_c(\beta)$ (as shown by the red horizontal line). In the inset, we zoom in in the region where $\rho - \rho_c(\beta) = O(1/\sqrt{V})$ and show schematically how the crossover occurs from the left (fluid) to the right (condensed), as obtained by taking the logarithm in Eq. (\ref{Z_crossover}).}\label{Fig_logZ}
\end{figure}

In the previous two subsections, we have seen that, 
to leading order for large $V$, the free energy per unit volume undergoes a freezing 
transition at $\rho = \rho_c(\beta)$. More precisely, for small $(\rho-\rho_c(\beta))$, one has from Eqs. (\ref{quadratic}) and (\ref{summary_free})
\bea \label{summary}
-f(\rho,\beta) &=& \lim_{V \to \infty} \frac{\ln Z(N,\beta)}{V}   \nonumber   \\
&=&
\begin{cases}
 &  \frac{\zeta(d+1)}{\beta^d}  - \frac{\beta^d}{\mk{2} \zeta(d-1)} (\rho_c(\beta)-\rho)^2  \quad, \quad \rho < \rho_c(\beta) \; ({\rm fluid}) \\
& \\
& \frac{\zeta(d+1)}{\beta^d} \quad, \quad \hspace*{3.3cm} \rho > \rho_c(\beta) \; ({\rm condensed}) \;.
\end{cases} 
\eea
Thus the second derivative of $f(\rho,\beta)$ with respect to $\rho$ is discontinuous at $\rho = \rho_c(\beta)$. It is interesting to understand how, for large but finite $V$, the free energy per unit volume of the two phases get continuously connected. In other words, if we zoom out the region in the vicinity of $\rho = \rho_c(\beta)$ in Fig. \ref{Fig_logZ} and appropriately scale $(\rho-\rho_c(\beta))$ with $V$, how do the left and right side of $\rho_c(\beta)$ in Fig. \ref{Fig_logZ} match smoothly with each other?

To achieve this matching, we start with the expression for the partition function in Eq. (\ref{ZN_sp5}), which is actually valid for both $\rho < 
\rho_c(\beta)$ and $\rho > \rho_c(\beta)$. Next we rescale $s \sqrt{V} = \tilde s$ to get
\bea
 \label{ZN_sp6}
Z(N,\beta) \approx e^{V \frac{\zeta(d+1)}{\beta^d}}\, \int_{\Gamma} \frac{d\tilde s}{2\pi i}\, \frac{1}{\tilde s} \, e^{\sqrt{V}\left( \rho - \rho_c(\beta)\right)\tilde s + \frac{\zeta(d-1)}{\mk{2}\beta^d}\,\tilde s^2 + \cdots} \;.
\eea 
We now consider the limit $\rho \to \rho_c(\beta)$ and $V \to \infty$ but with the product 
$\sqrt{V}(\rho-\rho_c(\beta)) = y$ fixed. In this limit, one can see that all the higher order terms denoted by 
$\cdots$ in (\ref{ZN_sp6}) drop out of the integral, leading to 
\bea \label{ZN_sp7}
Z(N,\beta) \approx e^{V \frac{\zeta(d+1)}{\beta^d}}\,  \int_{\Gamma} \frac{d\tilde s}{2\pi i}\, \frac{1}{\tilde s} \, 
e^{y\tilde s + \frac{\zeta(d-1)}{\mk{2}\beta^d}\,\tilde s^2} \;.
\eea
Taking a derivative with respect to $y$ makes it a simple Gaussian integral and one gets
\bea \label{deriv_Z}
\frac{\partial}{\partial y}\, Z(N,\beta) \approx e^{V \frac{\zeta(d+1)}{\beta^d}}\,\frac{1}{\sqrt{4 \pi b}}\,e^{-\frac{y^2}{4 b}} 
\quad, \quad b=  \frac{\zeta(d-1)}{\mk{2}\beta^d} \;.
\eea
\mk{Integrating this back with respect to $y$ and using the fact that $Z(N,\beta) \to 0$ as $y \to -\infty$ one gets a scaling form
\bea \label{Z_crossover}
Z(N,\beta) \approx e^{V \frac{\zeta(d+1)}{\beta^d}} \, F^{\rm normal}_{c}\left(\frac{(\rho-\rho_c(\beta))\sqrt{V}}{\sqrt{4 b}}\right)  \quad, \quad {\rm with} \quad b=  \frac{\zeta(d-1)}{\mk{2}\beta^d}
\eea
where the scaling function $F_c^{\rm normal}(z)$ connecting the two phases is given by
\bea \label{Fc_norm}
F_c^{\rm normal}(z)= \frac{1}{2} \left(1+{\rm erf}(z) \right) \;.
\eea}
Here ${\rm erf}(x) = 2/\sqrt{\pi} \int_0^x e^{-u^2} \,du$ is the error function and we used $y = (\rho-\rho_c(\beta))\sqrt{V}$ in Eq. (\ref{ZN_sp7}). We recall that $d>2$. The asymptotic behaviours of the error function read
\bea \label{asympt_erf}
{\rm erf}(x) \approx 1 - \frac{e^{-x^2}}{|x| \sqrt{\pi}} \quad, \quad x \to \pm \infty \;. 
\eea
Therefore, when $(\rho - \rho_c(\beta))>0$ and $(\rho - \rho_c(\beta))\gg 1/\sqrt{V}$ one gets from (\ref{Z_crossover}) and (\ref{asympt_erf}) for large positive $x$, one recovers the result in the condensed phase given in the second line of Eq. (\ref{summary}). In contrast, for $(\rho - \rho_c(\beta))<0$ and $(\rho_c(\beta) - \rho)\gg 1/\sqrt{V}$, one gets from (\ref{Z_crossover}) and (\ref{asympt_erf}) for large negative $x$ the behavior in the fluid phase given in the first line of Eq. (\ref{summary}). Thus the result in Eq. (\ref{Z_crossover}) is the crossover function that smoothly connects the partition functions on both sides. Let us remark that, to compute this crossover function accurately, we needed to perform the Bromwich integral close to $s=0$. If, instead, we had used the effective saddle point approach
mentioned before, it would have given the leading order $e^{O(V)}$ term in Eq. (\ref{Z_crossover}) correctly, but not the crossover 
function which requires a more careful analysis of the Browmwich integral, as done above.

\mk{Let us emphasize here that the crossover scaling function $F_c^{\rm norm}(z)$ in Eq. (\ref{Fc_norm}) is valid again for normal condensation, i.e.,
for $d>2$. For $1<d<2$ where the condensation transition is anomalous, one can
similarly compute the crossover function connecting the fluid and the condensed phases.
It is not difficult to show that in this case, i.e., for $1<d<2$, the analogue of Eq. (\ref{Fc_norm}) becomes \cite{HKMS}
\begin{equation}
Z(N, \beta)= e^{V \frac{\zeta(d+1)}{\beta^d}}\, F^{\rm anom}_c\left( \frac{(\rho-\rho_c(\beta))\, 
V^{1-1/d}}{c}\right)\quad {\rm with}\quad c= \frac{1}{\beta}\, \left(\frac{\Gamma(2-d)}{(d-1)d}\right)^{1/d}\, ,
\label{cross.1}
\end{equation}
where the crossover scaling function is given by
\begin{equation}
F^{\rm anom}_c(z)= \int_{\Gamma} \frac{d \tilde{s}}{2\pi i}\, \frac{1}{\tilde{s}}\, e^{z\, \tilde{s} +
{\tilde{s}}^d}\, .
\label{cross.2}
\end{equation}
Taking a derivative and then integrating back, one can express it as
\begin{equation}
F^{\rm anom}_c(z)= \int_{-z}^{\infty} V_{d+1}(z')\, dz'\, ,
\label{cross.3}
\end{equation}
where 
\begin{equation}
V_\gamma(z)= \int_{\Gamma} \frac{ds}{2\pi i} e^{-z\, s+ s^{\gamma-1}}\, ,
\label{vg.z}
\end{equation}
is precisely the scaling function that appeared in the analysis of anomalous condensate
in zero range processes \cite{MEZ05,EMZ06}. The detailed asymptotic properties of $V_\gamma(z)$ can
be found in Ref. \cite{EMZ06} [see Eqs. (77)-(83) and Fig. 5 for a plot therein]. Note that the function $V_\gamma(z)$ in Eq.~\eqref{vg.z} is normalized
to unity, i.e., $\int_{-\infty}^{\infty} V_\gamma(z) dz=1$.}

Let us end this section by drawing a comparison to a similar condensation transition observed in interacting particle systems, such as the 
zero range process (ZRP) \cite{EH05,SNM10}. The ZRP is defined on a lattice where each site contains $n_i$ particles and a single particle 
from site $i$ can jump to its neighboring sites with a rate that depends only on $n_i$. The dynamics conserves the total number of 
particles. In this model, for a certain choice of the hopping rate, the system undergoes a condensation phase transition when the density 
$\rho$ of particles exceeds a critical value. In the condensed phase, one single site carries a macroscopic fraction of particles. Thus, 
unlike in the Bose gas case studied here, where the condensation occurs in the momentum/energy space, in ZRP and related models the 
condensation occurs in the real space~\cite{MEZ05,EMZ06}. In the low density fluid phase, the grand canonical partition function can again 
be evaluated by the saddle point method and in this case the canonical and the grand-canonical ensembles are equivalent. However, in the 
condensed phase, the saddle hits the origin and, to evaluate the canonical partition function, one can no longer use the equivalence to the 
grand-canonical ensemble~\cite{MEZ05,EMZ06}. Instead, one has to evaluate the canonical partition by performing the Bromwich
 integral close to $s=0$~\cite{MEZ05,EMZ06}, as in the Bose 
gas case studied here. Similar real space condensation scenarios appeared recently in a number of contexts: this includes, e.g., the 
distribution of the position of a run-and-tumble particle in $d$-dimensions \cite{gradenigo2019,mori2021,mori2021b} and the 
localization/delocalization transition in discrete nonlinear Schr\"odinger equation~\cite{NLS1,NLS2}. However, there is an important 
difference in details between such real space condensation models like the ZRP and the noninteracting trapped Bose gas considered here. In 
the ZRP case, the canonical partition function can be expressed in a similar way as in Eq. (\ref{ZN_sp2}) for the Bose gas, with the 
important difference that the corresponding action $A(s,\rho)$ vanishes at $s=0$ in the ZRP~\cite{EMZ06}, while it is non zero in the Bose 
gas case. \mk{Hence the freezing transition that we find here is specific to the Bose gas and is absent in the mass transport models like ZRP.}

\section{Statistics of the moment of inertia for $d>2$} \label{sec:stat_I}

For simplicity, we again restrict ourselves to the case $d>2$. However, a similar analysis can be carried out also for $1<d<2$. In this section we study the statistics of $I = \omega^2 \sum_i \vec{r_i}^2$ at finite temperature $T$, where we have set 
$m=1$ in Eq.~(\ref{def_I}).  Having obtained the partition function $Z(N,\beta)$ in the large $N$ limit in the previous 
section, we will now use these results in the exact relation in Eq. (\ref{relation_tilde}) to compute the statistics of $I$.

We first start with the average $\langle I \rangle = -\partial_\beta \ln Z(N,\beta)$ given in Eq. (\ref{av_I_2}). In the fluid 
phase where $\rho<\rho_c(\beta)$, by using the expression of $\ln Z(N,\beta)/V$ in Eqs. (\ref{ZN_sp3}) and (\ref{quadratic}), 
and taking derivative with respect to (w.r.t.) $\beta$, one can in principle implicitly compute this average for all $\rho<\rho_c(\beta)$. 
However, it is easier to work with the expansion of $\ln Z(N,\beta)/V$ close to $\rho = \rho_c(\beta)$ in either phases, as 
given in Eq. (\ref{summary}). Taking a derivative of that expression in (\ref{summary}) w.r.t $\beta$ one gets, 
up to the linear order in $(\rho-\rho_c(\beta))$
\bea \label{av_I_final} 
\frac{\langle I \rangle}{V} \underset{V \to \infty}{=} 
\begin{cases}
& \mk{\frac{d\, \zeta(d+1)}{\beta^{d+1}} - \frac{d\, \zeta(d)}{ \zeta(d-1) \beta}(\rho_c(\beta)-\rho) + o(\rho_c(\beta)-\rho) \quad, \quad \hspace*{0.7cm} \rho < \rho_c(\beta)} \;, \\
& \\
& \frac{d\, \zeta(d+1)}{\beta^{d+1}} \quad, \quad \hspace*{6.7cm} \rho > \rho_c(\beta) \;.
\end{cases}
\eea
Thus while the average $\langle I\rangle/V$ {is itself continuous} at $\rho=\rho_c(\beta)$,
its first derivative w.r.t $\rho$ is discontinuous at $\rho=\rho_c(\beta)$ \mk{and furthermore, its value freezes for $\rho>\rho_c(\beta)$, i.e., it becomes independent of $\rho$ as $\rho$ increases beyond  $\rho_c(\beta)$.} {This is our first
real space diagnostic of the BEC transition.}

Let us now turn to the variance. Our starting point is the exact relation in Eq. (\ref{rel_var}), which holds for all $\omega$. However, since we will work in the appropriate thermodynamic limit where $\omega \to 0$ and $N \to \infty$ with $N\, \omega^d = \rho$ fixed, we first take the $\omega \to 0$ limit in Eq. (\ref{rel_var}) which then reads
\bea \label{var_smallom}
{\rm Var}(I) = \langle I^2 \rangle -  \langle I \rangle^2 \underset{\omega \to 0}{=} \left[ \frac{\partial^2 \ln Z}{\partial \beta^2} + \frac{1}{\beta} \langle I \rangle\right] \;.
\eea
We now take two derivatives w.r.t. $\beta$ in Eq. (\ref{summary}) and also use the expression of $\langle I \rangle$ in Eq. (\ref{av_I_final}). After a long but straightforward algebra we get


{\bea \label{varI_final}
\frac{{\rm Var}(I)}{V} \underset{V \to \infty}{=} 
\begin{cases}
\begin{aligned}
& \mk{\frac{1}{\beta^{d+2}}
  \!\left[
    d(d+2)\zeta(d+1)
    - \frac{d^2 \zeta^2(d)}{\, \zeta(d-1)}
  \right]
  \quad, \quad \text{for}\quad \rho \to \rho_c(\beta)^-} \;, \\[6pt]
& \frac{d(d+2)\, \zeta(d+1)}{\beta^{d+2}}
  \quad, \quad \;\; \;\;\; \hspace*{2.6cm} \text{for}\quad\rho > \rho_c(\beta) \;.
\end{aligned}
\end{cases}
\eea}
Thus, as $\rho$ approaches $\rho_c(\beta)$ from above and below, the variance of $I$ is discontinuous across 
$\rho = \rho_c(\beta)$. This discontinuity provides another real space signature of the BEC transition. \mk{In addition, as in the case of $\langle I \rangle$, the variance also freezes for $\rho>\rho_c(\beta).$}

Having obtained the nonanalytic behaviors of both the mean and the variance at the critical density $\rho=\rho_c(\beta)$,
it is natural to wonder how the full distribution of $I$ behave as one changes the density $\rho$.  \mk{It is natural to assume that, away from the critical density, the distribution $P_\beta(I,N)$ in Eq. (\ref{PDF_I}) admits a large deviation principle in the thermodynamic limit (when $\omega \to 0$, or equivalently $V = \omega^{-d} \to \infty$ and also $N \to \infty$ but with the 
density $\rho = N/V$ kept fixed) of the form}
\bea \label{LDF}
P_\beta(I,N) \approx e^{-V\, \Phi\left( \frac{I}{V}\right)} \;,
\eea
where $\Phi(z)$ is a rate function. Assuming that such a large deviation form holds, one can then relate the rate function $\Phi(z)$ to the cumulant generating function of $P_\beta(I,N)$. Indeed, substituting this anticipated form (\ref{LDF}) in the Laplace transform $\tilde P_\beta(\lambda,N)$ in Eq. (\ref{Laplace_PDF_I}) gives
\bea \label{LDF_Laplace}
\tilde P_\beta(\lambda,N) = \langle e^{-\lambda I} \rangle \approx \int_0^\infty e^{-\lambda I - V\, \Phi\left( \frac{I}{V}\right)}\, dI \;.
\eea
Performing the change of variable $I/V = z$ and evaluating the resulting integral over $z$ by a saddle point method, assuming it exists, one gets
\bea \label{LDF_Laplace2}
\tilde P_\beta(\lambda,N)  \approx e^{-V\, \Psi(\lambda)} \quad, \quad {\rm with} \quad \Psi({\lambda}) = \min_{z} \left[ \lambda z + \Phi(z) \right] \;.
\eea  
Taking logarithm on both sides, we see that 
\bea \label{CGF}
\Psi(\lambda) \approx - \frac{1}{V} \ln \, \langle e^{- \lambda I} \rangle
\eea
is nothing but the cumulant generating function of the random variable $I$. By inverting the Legendre transform in Eq. (\ref{LDF_Laplace2}), one can in principle express $\Phi(z)$ in terms of $\Psi(\lambda)$ via the relation
\bea \label{inv_Legendre}
\Phi(z) = \max_{\lambda} \left[ - \lambda\,z + \Psi(\lambda) \right] \;.
\eea
Thus, to derive $\Phi(z)$, we need to determine $\Psi(\lambda)$ and then Eq. (\ref{inv_Legendre}) gives access to $\Phi(z)$. We will see below that the rate function $\Phi(z)$ has two distinct behaviors respectively in the fluid phase ($\rho < \rho_c(\beta)$) and in the condensed phase ($\rho>\rho_c(\beta)$). 



\subsection{Computation of $\Psi(\lambda)$} \label{sub:psi}

\begin{figure}[t]
\centering
\includegraphics[width=0.6\linewidth]{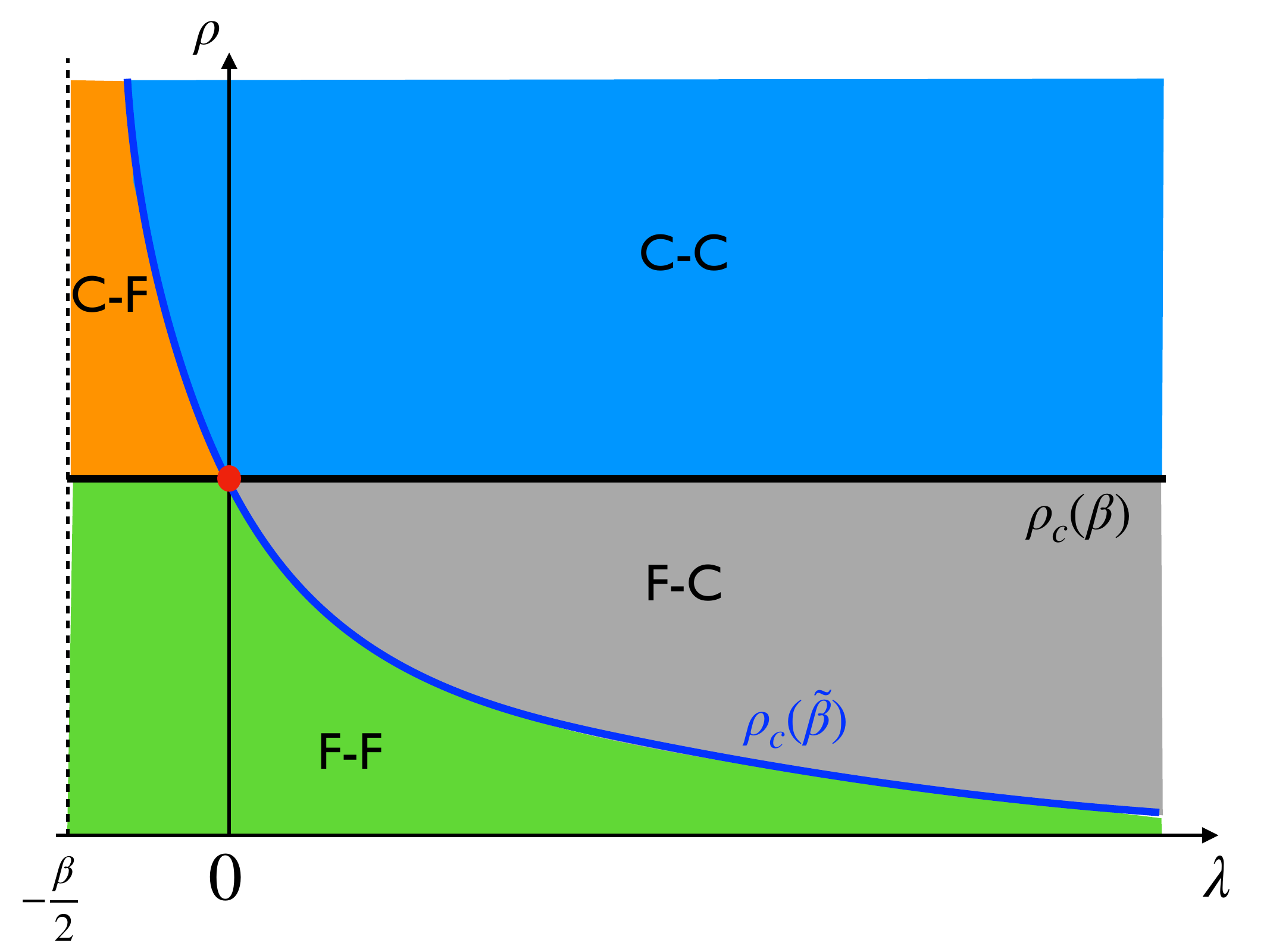}
\caption{Phase diagram in the $(\lambda,\rho)$ plane. The allowed values of $\lambda$ lie in the range $\lambda \in (-\beta/2,+\infty)$ and $\rho  \geq 0$. The black horizontal line represents the physical critical density $\rho = \rho_c(\beta)$, which is independent of $\lambda$. The solid blue curve represents $\rho_c(\tilde \beta)$ with $\tilde \beta = \sqrt{\beta(\beta + 2 \lambda)}$, as given in Eq. (\ref{rhoc_tilde}) and it diverges as $\lambda \to -\beta/2$. These two curves meet at $(\lambda=0, \rho = \rho_c(\beta))$, marked by a red filled circle. The four colored phases representing the fluid-fluid (F-F), the fluid-condensed (F-C), the condensed-fluid (C-F) and the condensed-condensed (C-C) meet at the red filled circle.}\label{Fig_phdiag}
\end{figure}

To determine $\Psi(\lambda)$, we can now make use of the exact relation in Eq. (\ref{relation_tilde}) and 
the already derived large $V$ behaviour of the partition function $Z(N,\beta)$ in the previous section.
To use this relation (\ref{relation_tilde}), we need to first determine $\tilde \beta(\lambda)$ as a function of $\beta$ and $\lambda$ using Eq. (\ref{rel_btilde}). Taking the $\omega \to 0$ limit (since we are working in the thermodynamic limit) gives simply
\bea  \label{btilde_expl}
\tilde \beta = \sqrt{\beta(\beta + 2 \lambda)} \;.
\eea
For $\tilde \beta$ to be real, we need the hard constraint $\lambda \geq - \beta/2$. This simply means that the Laplace transform $\langle e^{-\lambda I}\rangle$ in Eq. (\ref{LDF}) diverges as $\lambda \to -\beta/2$ indicating that $\lambda \in [-\beta/2, + \infty)$ is the allowed range of the real part of $\lambda$. 
Using this relation in Eq. (\ref{rel_btilde}) and (\ref{LDF_Laplace2}) then gives, in the thermodynamic limit, 
\bea  \label{rel_btilde_thermo}
\tilde P_\beta(\lambda,N) = \langle e^{-\lambda I} \rangle\approx e^{- V \Psi(\lambda)} \approx \frac{Z(N,\sqrt{\beta(\beta + 2 \lambda)})}{Z(N,\beta)} \;.
\eea
Taking logarithm on both sides then gives the desired rate function $\Psi(\lambda)$ in terms of the partition function $Z(N,\beta)$
\bea \label{Psi_Z}
\Psi(\lambda) = \lim_{V \to \infty} \left[ -\frac{1}{V} \ln Z\left(N,\sqrt{\beta(\beta + 2 \lambda)}\right) + \frac{1}{V} \ln Z(N,\beta)\right] \;.
\eea
Therefore $\Psi(\lambda)$ has a nice physical interpretation. Recalling the definition of the free energy per unit volume 
in Eq. (\ref{def_free_e}),
we see that $\Psi(\lambda)$ represents the free energy difference between two ``species'' of ideal Bose gases: first (A) which 
is held at inverse temperature $\beta$ and the second (B) which is held at inverse temperature $\tilde \beta = 
\sqrt{\beta(\beta + 2 \lambda)}$ given in Eq. (\ref{btilde_expl}). Thus Eq. (\ref{Psi_Z}) reads
\bea \label{Psi_free}
\Psi(\lambda) = f_{\rm B}\left(\rho,\beta\right) - f_{\rm A}(\rho,\beta) \;,
\eea
\mk{where we have defined
\bea
f_A(\rho,\beta) &=& - \lim_{V \to \infty} \frac{1}{V} \ln Z(N,\beta) \label{def_fA}\\ 
 f_B(\rho,\beta) &=& - \lim_{V \to \infty} \frac{1}{V} \ln Z\left(N,\tilde \beta = \sqrt{\beta(\beta + 2 \lambda)}\right) \;.\label{def_fB}
\eea}
For the species A, the critical density that separates its fluid phase from its condensed counterpart 
is $\rho_c(\beta)$ given in Eq. (\ref{rhoc}), which of course is independent of $\lambda$. For species B, on the other hand, the corresponding
critical density is given by $\rho_c(\tilde \beta)$ where $\tilde \beta$ is given in (\ref{btilde_expl}) and clearly it depends on $\lambda$. The critical densities for the two species are given explicitly by 
\bea \label{rhoc_tilde}
\rho_c(\beta) = \frac{\zeta(d)}{\beta^d} \quad, \quad \rho_c(\tilde \beta) = \frac{\zeta(d)}{\tilde \beta^d} = \frac{\zeta(d)}{[\beta(\beta + 2 \lambda)]^{d/2}} \;.
\eea
Thus, for a fixed inverse temperature $\beta$, the function $\Psi(\lambda)$ exhibits different behaviors in the 
$(\lambda,\rho)$ plane. We recall that the allowed range of $\lambda$ is $\lambda \in [-\beta/2, +\infty)$ and $\rho$ can be 
any positive number. This allowed range in the $(\lambda, \rho)$ plane is shown in Fig. \ref{Fig_phdiag}. In this figure, the 
horizontal black line denotes $\rho_c(\beta)$ for the species A given in (\ref{rhoc_tilde}) which, as mentioned before, is independent of 
$\lambda$. The dark blue line denotes the other critical density $\rho_c(\tilde \beta)$ in (\ref{rhoc_tilde}) for
the species B. These two 
critical lines divide the allowed region in $(\lambda,\rho)$ plane into four distinct phases (see Fig. \ref{Fig_phdiag}):
\begin{itemize}
\item[(i)]{The fluid-fluid (F-F) phase where $\rho<\rho_c(\beta)$ and $\rho< \rho_c(\tilde \beta)$. In this phase, both species A and B are in their respective fluid phases.}
\item[(ii)]{The fluid-condensed (F-C) phase where $\rho<\rho_c(\beta)$ but $\rho>\rho_c(\tilde \beta)$. Here, the species A is still in its fluid phase but species B is in its condensed phase.}
\item[(iii)]{The condensed-fluid (C-F) phase where $\rho>\rho_c(\beta)$, while $\rho< \rho_c(\tilde \beta)$. In this case, the species A is in the condensed phase, while the species B is in the fluid phase.}
\item[(iv)]{The condensed-condensed (C-C) phase where $\rho>\rho_c(\beta)$ and $\rho> \rho_c(\tilde \beta)$. Here both species are in their condensed phases.}
\end{itemize}
Since the partition function for each species behaves differently across its critical density, the free energy per unite 
volume $f_{\rm A}(\rho,\beta)$ and $f_{\rm B}(\rho,\beta)$ will also behave differently across the critical densities for each 
species. Hence, $\Psi(\lambda)$ in Eq. (\ref{Psi_free}) will also behave differently in each of the four different phases in 
the $(\lambda, \rho)$ plane in Fig. \ref{Fig_phdiag}. These different behaviors are derived below. For convenience, we will 
denote the free energies per unit volume for each species A and B, respectively in the fluid and the condensed phases as 
$f_{{\rm A,B}}^{\rm F}(\rho,\beta)$ and $f_{{\rm A,B}}^{\rm C}(\rho,\beta)$, where the superscripts F and C refer to the fluid 
and the condensed phases.

\subsubsection{The F-F phase}

\begin{figure}[t]
\centering
\includegraphics[width = 0.9 \linewidth]{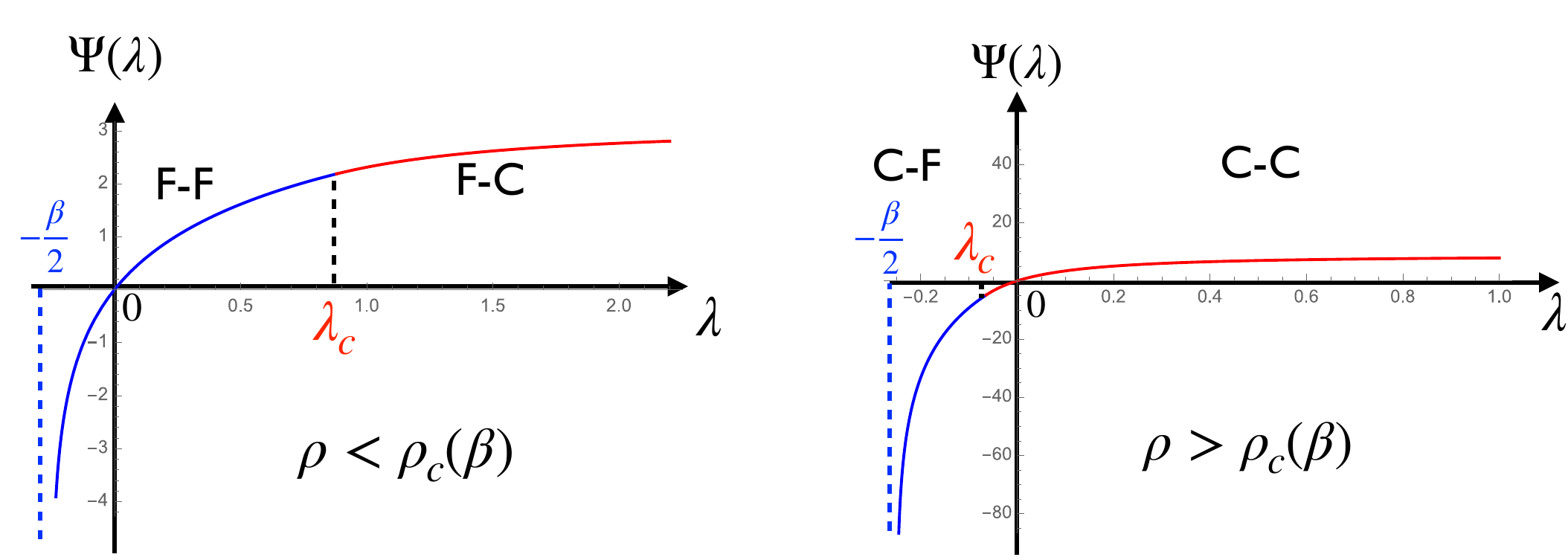}
\caption{The exact expression of the function $\Psi(\lambda)$ in the large $V$ limit, as detailed in the text,  plotted as a function of $\lambda$ for $\rho<\rho_c(\beta)$ (left panel) and $\rho>\rho_c(\beta)$ (right panel). In both panels, $\lambda_c$ marks the value of $\lambda$ that separates two phases as one increases $\lambda$ and the value of $\lambda_c$ is given in Eq. (\ref{lmax}). In the figure, we used $\beta = 1/2$. On the left panel, $\lambda_c$ separates the F-F and the F-C phases, while on the right panel it separates the C-F and the C-C phases. As $\lambda$ crosses $\lambda_c$, in both panels, the functions $\Psi(\lambda)$ and its first derivative are continuous, while the second derivative  $\Psi''(\lambda)$ undergoes a discontinuous jump (though not visible in the figure). \mk{Here we used $d=3$}.}\label{Fig_Psi}
\end{figure}

In this phase both species are in their respective fluid phases, since $\rho < \rho_c(\beta)$ and $\rho< \rho_c(\tilde \beta)$.  Hence, in this case, we have 
\bea \label{Psi_FF0}
\Psi(\lambda) = f_{\rm B}^{\rm F}(\rho,\beta) - f_{\rm A}^{\rm F}(\rho,\beta) \;.
\eea
We can then use the expression given in Eq. (\ref{ZN_sp3}) for both species. This leads to
\bea \label{Psi_FF}
\Psi(\lambda) = - \rho \tilde s^* - \frac{1}{\tilde \beta^d}\, {\rm Li}_{d+1}(e^{-\tilde s^*}) + \rho s^* + \frac{1}{\beta^d}\, {\rm Li}_{d+1}(e^{-s^*})
\eea
where $s^*$ and $\tilde s^*$ denote respectively the saddle points for species A and B and are given implicitly as the solutions of 
\bea \label{sstilde}
{\rm Li}_d(e^{-s^*}) = \rho\, \beta^d \quad, \quad {\rm and} \quad {\rm Li}_d(e^{-\tilde s^*}) = \rho\, \tilde \beta^d = \rho\, [\beta(\beta+2\lambda)]^{d/2} \;.
\eea
We compute $\Psi(\lambda)$ for  a fixed $\rho$ in the F-F phase by varying $\lambda$. 
For a given $\rho$ in this phase, we note that the condition $\rho<\rho_c(\tilde \beta)$ indicates that $\lambda < \lambda_{c}$ where 
\bea  \label{lmax}
 \lambda_{c}= \frac{1}{2} \left(\frac{1}{\beta} \left[\frac{\zeta(d)}{\rho} \right]^{2/d} - \beta\right) \;.
\eea
Therefore in this phase, the allowed range of $\lambda$ is $-\beta/2 < \lambda < \lambda_{c}$. In principle, for a given $\rho$ and $\lambda$ in the F-F phase, one can obtain $s^*$ and $\tilde s^*$ numerically from Eq. (\ref{sstilde}) and, hence, determine $\Psi(\lambda)$ numerically from Eq. (\ref{Psi_FF}). A plot of this $\Psi(\lambda)$ is given in Fig. \ref{Fig_Psi}. However, one can extract analytically the asymptotic behavior of $\Psi(\lambda)$ as $\lambda \to -\beta/2$ as well as $\lambda \to \lambda_{c}$.

\vspace*{0.5cm}
\noindent{\bf The limit $\lambda \to - \beta/2$.} In this case $\tilde \beta \to 0$ from Eq. (\ref{btilde_expl}). From Eq. (\ref{sstilde}), together with the asymptotic expansion given in Eq. (\ref{h_asympt}), it follows that, to leading order, $e^{-\tilde s^*} \approx \rho \tilde \beta^d$. Consequently, 
\bea \label{sstar_left}
\tilde s^* \to -\frac{d}{2}\, \ln\left(\beta + 2 \lambda\right) \quad, \quad {\rm as} \quad \lambda \to -\beta/2 \;.
\eea 
Substituting this behavior back in Eq. (\ref{Psi_FF}) one finds, to leading order
\bea \label{Psi_left}
\Psi(\lambda) \approx  \rho \frac{d}{2} \ln \left(\beta + 2 \lambda \right) + c_0  \quad, \quad {\rm as} \quad \lambda \to -\beta/2  \;,
\eea
where $c_0$ is an unimportant constant independent of $\lambda$.

\vspace*{0.5cm}
\noindent{\bf The limit $\lambda \to \lambda_{c}$.} In this limit, the species B is close to its critical density $\rho_c(\tilde \beta)$. Hence we can use the saddle point approximation of the free energy given in the first line of Eq. (\ref{summary}), after replacing $\beta$ by $\tilde \beta$. 
This gives
\bea \label{fB_FF}
-f^{\rm F}_{\rm B}(\rho,\beta) = \frac{\zeta(d+1)}{[\tilde \beta(\lambda)]^d} - \frac{[\tilde \beta(\lambda)]^d}{\mk{2} \zeta(d-1)} \left(\frac{\zeta(d)}{[\tilde \beta(\lambda)]^d} -\rho \right)^2 \;,
\eea 
where we used $\rho_c(\tilde \beta) = \zeta(d)/[\tilde \beta(\lambda)]^d$ and we recall that $\tilde \beta(\lambda) = \sqrt{\beta(\beta+2\lambda)}$. We now need to expand $f_{\rm B}(\rho,\beta)$ in Eq.~(\ref{fB_FF}) for $\lambda$ close to $\lambda_{c}$ where $\lambda_{c}$ is given in Eq. (\ref{lmax}). We set $\lambda = \lambda_{c} - \epsilon$ in the expression for $\tilde \beta(\lambda)$ in Eq. (\ref{btilde_expl}). We also notice from Eq. (\ref{lmax}) that
\bea \label{lmax2}
[\beta(\beta + 2 \lambda_{c})]^{d/2} = \frac{\zeta(d)}{\rho} \;.
\eea
By expanding up to order $O(\epsilon^2)$ we then get
\bea \label{exp_betat}
[\tilde \beta(\lambda)]^d = \frac{\zeta(d)}{\rho} \left[1-  \frac{d}{(\beta+2 \lambda_{c})}\epsilon + \frac{d(d-2)}{2(\beta+2\lambda_{c})^2}\epsilon^2 + O (\epsilon^3)\right] \;.
\eea
Substituting this expansion in the expression of $f^{\rm F}_{\rm B}(\rho,\beta) $ in Eq. (\ref{fB_FF}) we get
\bea \label{fB_approx}
f^{\rm F}_{\rm B}(\rho,\beta) = - \frac{\rho \zeta(d+1)}{\zeta(d)} - \frac{d \zeta(d+1)}{\zeta(d)} \frac{\rho}{\beta + 2 \lambda_{c}} \, \epsilon - A_2\,\epsilon^2 + O(\epsilon^3) \;,
\eea
where
\bea \label{A2}
A_2= \frac{ d\,\rho\,\left(-d [\zeta(d)]^2 + (d+2)\zeta(d-1)\zeta(d+1) \right)}{2 \zeta(d)\zeta(d-1)(\beta + 2 \lambda_{c})^2} \;.
\eea
One can check that $A_2 > 0$. 

On the other hand, for the species A, its free energy is independent of $\lambda$ and corresponds to its fluid phase and hence is given by
$f_{\rm A}^{\rm F}(\rho,\beta)$. Therefore $\Psi(\lambda)$ in Eq. (\ref{Psi_free}), close to $\lambda_{c}$ in the F-F phase behaves as
\begin{equation} \label{Psi_lmax}
\Psi(\lambda) = -f^{\rm F}_{\rm A}(\rho,\beta) - \frac{\rho \zeta(d+1)}{\zeta(d)} - \frac{d \zeta(d+1)}{\zeta(d)} \frac{\rho}{\beta + 2 \lambda_{c}} \, (\lambda_{c}-\lambda) - A_2\,(\lambda_{c}-\lambda)^2 + O((\lambda_{c}-\lambda)^3) \;.
\end{equation}
We will see later that the second derivative of $\Psi(\lambda)$, i.e., the amplitude of $(\lambda_{c}-\lambda)^2$ in this expansion, undergoes a discontinuous jump when $\lambda_{c}$ is approached from above, i.e., from the F-C side in Fig. \ref{Fig_phdiag}.

\subsubsection{The F-C phase}

Now we consider the F-C phase with $\rho< \rho_c(\beta)$ and $\rho> \rho_c(\tilde \beta)$. This corresponds to the grey shaded area in Fig.~\ref{Fig_phdiag}. We fix $\rho$ and vary $\lambda$. Clearly in this regime $\lambda$ varies from $\lambda_c$ to $+\infty$. We want to calculate $\Psi(\lambda)$ in this phase, for $\lambda \in [\lambda_c,+\infty)$. Noting that, for the species B, the system is in the condensed phase, the free energy $f^{\rm F}_{\rm B}(\rho,\beta)$ has a simple expression given in the second line of Eq. (\ref{summary_free}). Consequently, for any $\lambda \in [\lambda_c,+\infty)$, we get
\bea \label{Psi_FC}
\Psi(\lambda) = f^{\rm C}_{\rm B}(\rho,\beta) - f^{\rm F}_{\rm A}(\rho,\beta) = - f^{\rm F}_{\rm A}(\rho,\beta)  - \frac{\zeta(d+1)}{[\tilde \beta(\lambda)]^d}
= - f^{\rm F}_{\rm A}(\rho,\beta) - \frac{\zeta(d+1)}{[\beta(\beta+2\lambda)]^{d/2}} \;.
\eea
We recall that $f^{\rm F}_{\rm A}(\rho,\beta)$ is independent of $\lambda$. Below, we will compute the asymptotic behaviors of $\Psi(\lambda)$ in the two opposite limits $\lambda \to \infty$ and $\lambda \to \lambda_c$ (from above, i .e., $\lambda > \lambda_c$). 

\vspace*{0.5cm}
\noindent{\bf The limit $\lambda \to +\infty$.} In this limit, it follows trivially from Eq. (\ref{Psi_FC}) that
\bea \label{Psi_largel}
\Psi(\lambda) \approx - f^{\rm F}_{\rm A}(\rho,\beta) - \frac{\zeta(d+1)}{(2\beta\,\lambda)^{d/2}}  \quad, \quad {\rm as} \quad \lambda \to +\infty \;.
\eea

\vspace*{0.5cm}
\noindent{\bf The limit $\lambda \to \lambda_c$.} In this limit, substituting the expansion of $[\tilde \beta(\lambda)]^d$ from Eq. (\ref{exp_betat}) in Eq. (\ref{Psi_FC}), we get
\begin{equation} \label{Psi_lmax_plus}
\Psi(\lambda) = -f^{\rm F}_{\rm A}(\rho,\beta) - \frac{\rho \zeta(d+1)}{\zeta(d)} - \frac{d \zeta(d+1)}{\zeta(d)} \frac{\rho}{\beta + 2 \lambda_{c}} \, (\lambda_{c}-\lambda) - A'_2\,(\lambda_{c}-\lambda)^2 + O((\lambda_{c}-\lambda)^3) \;,
\end{equation}
where 
\bea \label{A2_prime}
A_2' = \frac{d\,(d+2)\,\rho  \, \zeta(d+1)}{2 \zeta(d)(\beta + 2 \lambda_c)^2} \;.
\eea 
Summarising the behavior of $\Psi(\lambda)$ on both sides close to $\lambda_c$ from Eqs. (\ref{Psi_lmax}) and (\ref{Psi_lmax_plus}), we find
\bea \label{psi_lc_fluid}
\Psi(\lambda) \approx
\begin{cases}
& -f^{\rm F}_{\rm A}(\rho,\beta) - \frac{\rho \zeta(d+1)}{\zeta(d)} - \frac{d \zeta(d+1)}{\zeta(d)} \frac{\rho}{\beta + 2 \lambda_{c}} \, (\lambda_{c}-\lambda) - A_2\,(\lambda_{c}-\lambda)^2  \;, \; \lambda \to \lambda_c^- \\
& \\
& -f^{\rm F}_{\rm A}(\rho,\beta) - \frac{\rho \zeta(d+1)}{\zeta(d)} - \frac{d \zeta(d+1)}{\zeta(d)} \frac{\rho}{\beta + 2 \lambda_{c}} \, (\lambda_{c}-\lambda) -A'_2\,(\lambda_{c}-\lambda)^2 \;, \; \lambda \to \lambda_c^+ \;,
\end{cases}
\eea 
Thus, comparing the expansions of $\Psi(\lambda)$ from above and below $\lambda_c$, we see that, while $\Psi(\lambda)$ and its first derivative $\Psi'(\lambda)$ are both continuous at $\lambda = \lambda_c$, the second derivative undergoes a discontinuous jump given by
\bea \label{jump_psi_sec}
\Psi''(\lambda_c^+) - \Psi''(\lambda_c^-) = 2 (A_2 - A'_2) =   -\frac{ \rho\,d^2 \zeta(d)}{\zeta(d-1)(\beta + 2 \lambda_{c})^2} =  -\frac{\beta^2\,d^2\, \rho^{\frac{4}{d}+1}}{ \zeta(d-1)[\zeta(d)]^{\frac{4}{d}-1}} \;,
\eea 
where in the last equality we eliminated $(\beta + 2 \lambda_c)$ using Eq. (\ref{lmax2}).
 
\subsubsection{The C-F and the C-C phases} 
 
The C-F phase corresponds to $\rho>\rho_c(\beta)$ and $\rho<\rho_c(\tilde \beta)$, while the C-C phase corresponds to $\rho>\rho_c(\beta)$ and $\rho>\rho_c(\tilde \beta)$. Thus, fixing $\rho > \rho_c(\beta)$, as we vary $\lambda$, the C-F phase corresponds to the region $\lambda \in (-\beta/2,\lambda_c]$ (shown by the orange shade in Fig. \ref{Fig_phdiag}) and the C-C phase corresponds to $\lambda \in [\lambda_c,+\infty)$ (shown by the blue shade in Fig. \ref{Fig_phdiag}). Here $\lambda_c$ is given by the same expression as in Eq. (\ref{lmax}). As $\lambda$ increases through $\lambda_c$, for fixed $\rho > \rho_c(\beta)$, the behaviour of $\Psi(\lambda)$ is rather similar to the analysis presented in the previous subsection for $\rho < \rho_c(\beta)$. The only difference is that the free energy per unit volume of the species A is now given by $f_{\rm A}^{\rm C}(\rho,\beta)$. More precisely $\Psi(\lambda)$ is given by
\bea \label{Psi_CF_CC}
\Psi(\lambda) = 
\begin{cases}
& -f_{\rm A}^{\rm C}(\rho,\beta) + f_{\rm B}^{\rm F}(\rho,\beta) \quad, \quad ({\rm C}-{\rm F} \quad {\rm phase}) \;,\\
& \\
& -f_{\rm A}^{\rm C}(\rho,\beta) + f_{\rm B}^{\rm C}(\rho,\beta) \quad, \quad ({\rm C}-{\rm C} \quad {\rm phase}) \;.
\end{cases}
\eea
For the contribution $f_{\rm B}^{\rm F}(\rho,\beta)$ and  $f_{\rm B}^{\rm C}(\rho,\beta)$ from the species B, one gets exactly the same expressions as in Eq. (\ref{summary_free}), with $\beta$ replaced by $\tilde \beta$. Skipping further details, let us summarise the main limiting behaviors of $\Psi(\lambda)$ as $\lambda$ varies from $-\beta/2$ to $+ \infty$. 

\vspace*{0.5cm}
\noindent{\bf The limit $\lambda \to - \beta/2$.} In this case 
\bea \label{Psi_left_2}
\Psi(\lambda) \approx  \rho \frac{d}{2} \ln \left(\beta + 2 \lambda \right) + c'_0  \quad, \quad {\rm as} \quad \lambda \to -\beta/2  \;,
\eea
where $c'_0$ is an unimportant constant independent of $\lambda$ and is different from $c_0$ in Eq. (\ref{Psi_left}). 

\vspace*{0.5cm}
\noindent{\bf The limit $\lambda \to \lambda_c$.} In this case, up to quadratic order in $(\lambda_c-\lambda)$ one gets
\bea \label{psi_lc}
\Psi(\lambda) \approx
\begin{cases}
& -f^{\rm C}_{\rm A}(\rho,\beta) - \frac{\rho \zeta(d+1)}{\zeta(d)} - \frac{d \zeta(d+1)}{\zeta(d)} \frac{\rho}{\beta + 2 \lambda_{c}} \, (\lambda_{c}-\lambda) - A_2\,(\lambda_{c}-\lambda)^2  \;, \; \lambda \to \lambda_c^- \\
& \\
& -f^{\rm C}_{\rm A}(\rho,\beta) - \frac{\rho \zeta(d+1)}{\zeta(d)} - \frac{d \zeta(d+1)}{\zeta(d)} \frac{\rho}{\beta + 2 \lambda_{c}} \, (\lambda_{c}-\lambda) - A'_2\,(\lambda_{c}-\lambda)^2 \;, \; \lambda \to \lambda_c^+ \;,
\end{cases}
\eea 
 where the constants $A_2$ and $A_2'$ are exactly the same as in Eqs. (\ref{A2}) and (\ref{A2_prime}) respectively. Thus, as in the case $\rho<\rho_c(\beta)$, the second derivative $\Psi''(\lambda)$ undergoes exactly the same discontinuous jump given in Eq. (\ref{jump_psi_sec}). 

\vspace*{0.5cm}
\noindent{\bf The limit $\lambda \to +\infty$.} In this case, the expression for $\Psi(\lambda)$ is given by
\bea \label{Psi_CC}
\Psi(\lambda) = f^{\rm C}_{\rm B}(\rho,\beta) - f^{\rm C}_{\rm A}(\rho,\beta) = - f^{\rm C}_{\rm A}(\rho,\beta)  - \frac{\zeta(d+1)}{[\tilde \beta(\lambda)]^d}
= - f^{\rm C}_{\rm A}(\rho,\beta) - \frac{\zeta(d+1)}{[\beta(\beta+2\lambda)]^{d/2}} \;.
\eea
Hence, for large $\lambda$, one gets 
\bea \label{Psi_largel2}
\Psi(\lambda) \approx - f^{\rm C}_{\rm A}(\rho,\beta) - \frac{\zeta(d+1)}{(2\beta\,\lambda)^{d/2}}  \quad, \quad {\rm as} \quad \lambda \to +\infty \;.
\eea

\subsection{Computation of the rate function $\Phi(z)$}\label{sub:phi}

\begin{figure}[t]
\centering
\includegraphics[width = 0.9\linewidth]{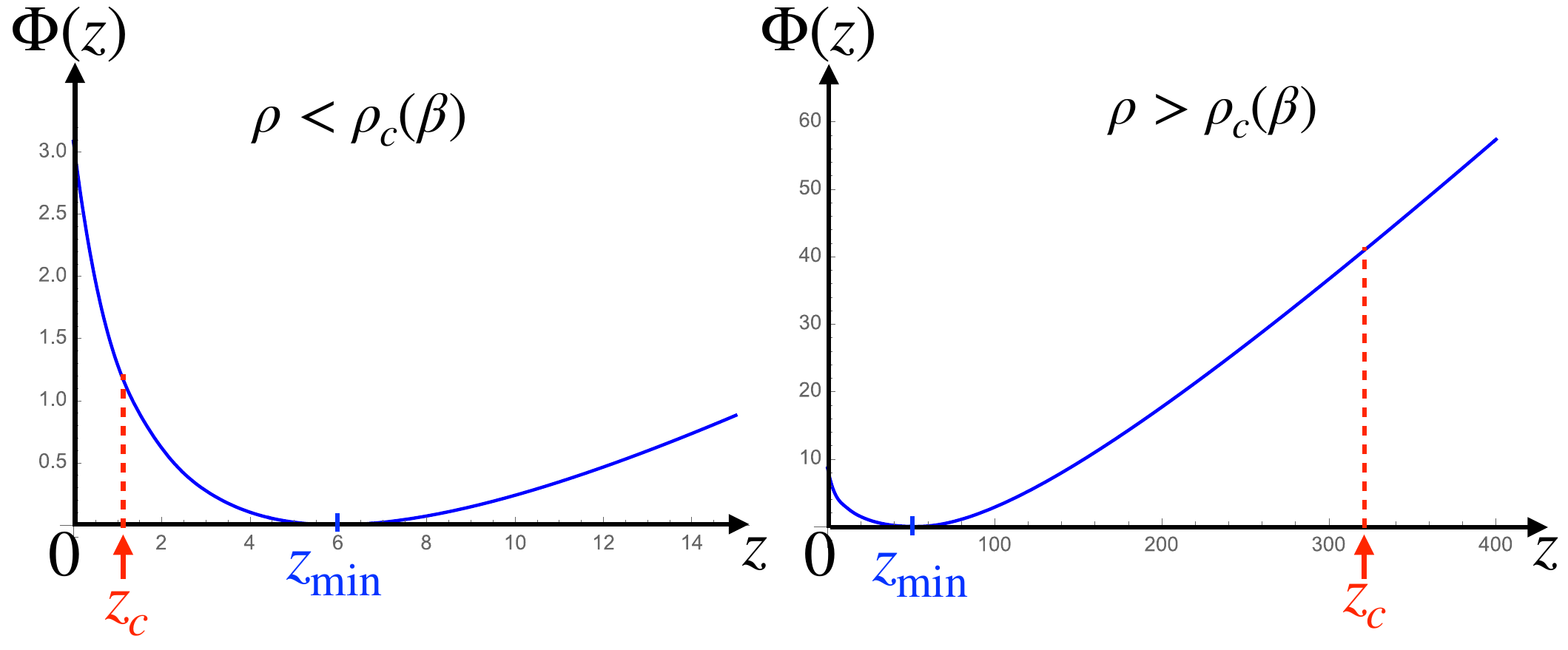}
\caption{Plot of the rate function $\Phi(z)$ vs $z$ for $\rho<\rho_c(\beta)$ (left panel) and $\rho>\rho_c(\beta)$ (right 
panel), which has been obtained by evaluating numerically the Legendre transform of $\Psi(\lambda)$ in Eq. (\ref{inv_Legendre2}), \mk{where for $\Psi(\lambda)$ we used its exact expressions in the large $V$ limit, depending on the four different phases, as detailed in Section \ref{sub:psi}}. In both panels, the function $\Phi(z)$ vanishes at $z_{\min}$ quadratically as in Eq. (\ref{Phi_quadratic}). The rate 
function $\Phi(z)$ has a singularity at $z=z_c$, given in Eq. (\ref{zc}). At $z=z_c$, while the functions $\Phi(z)$ and its 
first derivative are continuous, its second derivative undergoes a discontinuous jump as given in (\ref{jump_phi}). On the 
left panel where $\rho<\rho_c(\beta)$ (fluid phase), the singular point $z_c$ is to the left of $z_{\min}$, while on the right 
panel (condensed phase), $z_c$ is on the right side of $z_{\min}$. Here we used $\beta = 1/2$ \mk{and $d=3$}.}\label{Fig_rate}
\end{figure}

We recall that, provided the distribution $P_\beta(I,N)$ admits a large deviation form as in Eq. (\ref{LDF}), the associated rate function $\Phi(z)$ is related to $\Psi(\lambda)$ via the Legendre transform 
\bea \label{inv_Legendre2}
\Phi(z) = \max_{\lambda} \left[ - \lambda\,z + \Psi(\lambda) \right] \;.
\eea
In the previous section we have derived the behaviours of $\Psi(\lambda)$, as a function of $\lambda$, for fixed $\rho$ and $\beta$. In this section, we will use these expressions for $\Psi(\lambda)$ in Eq. (\ref{inv_Legendre2}) to determine the rate function $\Phi(z)$. One can in principle determine numerically $\Phi(z)$ for all $z$ from the known expression of $\Psi(\lambda)$. However, it is possible to extract analytically the asymptotic behaviors of $\Phi(z)$ in different limits, as shown below.

\subsubsection{The limit $z \to \infty$}

It turns out (and checked a posteriori) that, to determine the behaviour of $\Phi(z)$ as $z\to \infty$, we need to use the behavior of $\Psi(\lambda)$ as $\lambda$ approaches its lowest allowed value $\lambda \to -\beta/2$. Indeed, substituting the leading behaviour of $\Psi(\lambda)$ from Eqs. (\ref{Psi_left}) and (\ref{Psi_left_2}) as $\lambda \to -\beta/2$ in Eq. (\ref{inv_Legendre2}), we get
\bea \label{Phi_large_z1}
\Phi(z) \approx  \max_{\lambda} \left[ - \lambda\,z + \rho \frac{d}{2}\, \ln{(\beta + 2 \lambda)} \right]  \;,
\eea
where we have dropped the constant terms $c_0$ and $c'_0$ as they lead only to subleading corrections. The function $S(\lambda) =  - \lambda\,z + \rho \frac{d}{2}\, \ln{(\beta + 2 \lambda)}$ has a maximum at $\lambda_{\max}$ given by
\bea \label{lmax_largez}
\beta + 2\lambda_{\max} = \frac{\rho\,d}{z} \;.
\eea
We see here, a posteriori, that a small value of $\beta+ 2\lambda_{\max}$ indeed corresponds to large values of $z$. 
Substituting this $\lambda_{\max}$ in (\ref{Phi_large_z1}) gives
\bea \label{Phi_large_z2}
\Phi(z) = S(\lambda_{\max}) \approx \frac{\beta}{2}\,z - \frac{\rho\,d}{2}\, \ln z + O(1) \quad, \quad {\rm as\quad} z \to \infty \;.
\eea  

\subsubsection{The limit $z \to 0$}

In this limit, it turns out that we need to analyse the large $\lambda$ behavior of $\Psi(\lambda)$. Substituting the large $\lambda$ behaviours from Eqs. (\ref{Psi_largel}) and (\ref{Psi_largel2}) in (\ref{inv_Legendre2}) we get
\bea \label{Phi_smallz}
\Phi(z) = C + \max_{\lambda} \left[ -\lambda\,z - \frac{\zeta(d+1)}{(2 \beta \, \lambda)^{d/2}}\right]  \;,
\eea
where the constant $C$ (independent of $z$) is given by $C = -f_{\rm A}^{\rm F}(\rho,\beta)$ (for $\rho < \rho_c(\beta)$) and by $C = -f_{\rm A}^{\rm C}(\rho,\beta)$ (for $\rho > \rho_c(\beta)$). Performing this maximisation w.r.t. $\lambda$, it turns out that $\lambda_{\max} = A_0\,z^{-2/(d+2)}$ which clearly shows that $z$ must be small in order that $\lambda_{\max}$ is large, justifying a posteriori the use of the large $\lambda$ expansion of $\Psi(\lambda)$ in Eq. (\ref{Phi_smallz}). The constant $A_0$ is given by 
\bea \label{A0}
A_0 = \left( \frac{d}{2} \frac{\zeta(d+1)}{(2\beta)^{d/2}}\right)^{2/(d+2)} \;.
\eea
Using this $\lambda_{\max}$ in (\ref{Phi_smallz}) gives, for small $z$,
\bea \label{Phi_smallz2}
\Phi(z) \approx C - B_0 \, z^{\frac{d}{d+2}} \quad, \quad {\rm as} \quad z \to 0 \,
\eea
where the constant $B_0$ is simply related to $A_0$.

\subsubsection{The limit $z \to z_{\min}$}

We now consider the behavior of $\Phi(z)$ close to $z = z_{\min}$ where $z_{\min}$ denotes the value of $z$ at which $\Phi(z)$ has a global minimum. We will see that this amounts to analysing $\Psi(\lambda)$ close to $\lambda = 0$. From Eq. (\ref{CGF}), we see that the cumulant generating function $\Psi(\lambda)$ can always be expanded in a smooth power series of $\lambda$ for small $\lambda$
\bea \label{small_l}
\Psi(\lambda) = \frac{\langle I \rangle}{V}\,\lambda - \frac{{\rm Var}(I)}{2 V}\, \lambda^2 + O(\lambda^3) \;.
\eea
Thus, substituting this expansion up to the quadratic order in Eq. (\ref{inv_Legendre2}), we get 
\bea
\Phi(z) = \max_\lambda \left[- \left(z- \frac{\langle I \rangle}{V}\right) \lambda - \frac{1}{V} {\rm Var}(I)\,\lambda^2 \right] \;.
\eea
Maximising w.r.t. $\lambda$, one obtains quite generically a quadratic form for $\Phi(z)$ close to $z = z_{\min} = \langle I\rangle/V$ given by
\bea \label{Phi_quadratic}
\Phi(z) \approx \frac{\left(z - z_{\min} \right)^2}{2 \sigma^2} \quad, \quad {\rm where} \quad z_{\min} = \frac{\langle I \rangle}{V} \quad {\rm and} \quad \sigma^2 = \frac{{\rm Var}(I)}{V} \;.
\eea
Note that $z_{\min}$ and $\sigma^2$, in the $V \to \infty$ limit, can in principle be computed for any $(\rho, \beta)$ from the partition function via Eqs. (\ref{av_I_2}) and (\ref{var_smallom}). In general, they have quite complicated expressions but, near $\rho = \rho_c(\beta)$, one can compute them explicitly as given in Eqs. (\ref{av_I_final}) and (\ref{varI_final}).

\subsubsection{The limit $z \to z_c$}

It turns out that there is a ``critical'' point $z=z_c$ where the rate function $\Phi(z)$ and its derivative $\Phi'(z)$ are continuous, but the second derivative $\Phi''(z)$ is discontinuous. We will see that this actually corresponds to the singularity of $\Psi(\lambda)$ at $\lambda = \lambda_c$ found before. Indeed, substituting the expansions of $\Psi(\lambda)$ near $\lambda = \lambda_c$ from Eqs. (\ref{psi_lc_fluid}) and (\ref{psi_lc}), respectively for $\rho<\rho_c(\beta)$ and $\rho>\rho_c(\beta)$ in (\ref{inv_Legendre2}), and upon maximising w.r.t. $\lambda$ we find the following results. For $\rho<\rho_c(\beta)$, we get
\bea \label{phiz_zc_fluid}
\Phi(z) \approx
\begin{cases}
& f_{\rm A}^{\rm F}(\rho,\beta) - \frac{\rho\,\zeta(d+1)}{\zeta(d)} - \lambda_c\,z_c - \lambda_c(z-z_c) + \frac{(z-z_c)^2}{4 A_2} \quad, \quad z > z_c \\
& \\
& f_{\rm A}^{\rm F}(\rho,\beta) - \frac{\rho\,\zeta(d+1)}{\zeta(d)} - \lambda_c\,z_c - \lambda_c(z-z_c) + \frac{(z-z_c)^2}{4 A'_2} \quad, \quad z < z_c \;,
\end{cases}
\eea
where $z_c$ is given by
\bea \label{zc}
z_c = \Psi'(\lambda_c) = \beta\,d\,\zeta(d+1)\left( \frac{\rho}{\zeta(d)}\right)^{\frac{d+2}{d}} \;.
\eea
For $\rho>\rho_c(\beta)$, exactly the same behavior occurs, except that $f_{\rm A}^{\rm F}(\rho,\beta)$ in both lines of (\ref{phiz_zc_fluid}) \mk{gets replaced by  $f_{\rm A}^{\rm C}(\rho,\beta)$.} From this expression (\ref{phiz_zc_fluid}) we see that, while $\Phi(z)$ and $\Phi'(z)$ are continuous at $z=z_c$, the second derivative $\Phi''(z)$ undergoes a discontinuous jump given by
\bea \label{jump_phi}
\Phi''(z_c^+) - \Phi''(z_c^-) = \frac{1}{2} \left( \frac{1}{A_2} - \frac{1}{A'_2}\right) \;,
\eea
where $A_2$ and $A'_2$ are given explicitly in Eqs. (\ref{A2}) and (\ref{A2_prime}) respectively. Since $A'_2 > A_2$, the jump is nonnegative. Thus, at this critical point $z_c$, the rate function $\Phi(z)$ is nonanalytic with a discontinuous second derivative. Hence, this corresponds to a second order phase transition. 

\begin{figure}[ht]
\centering
\includegraphics[width=0.6\linewidth]{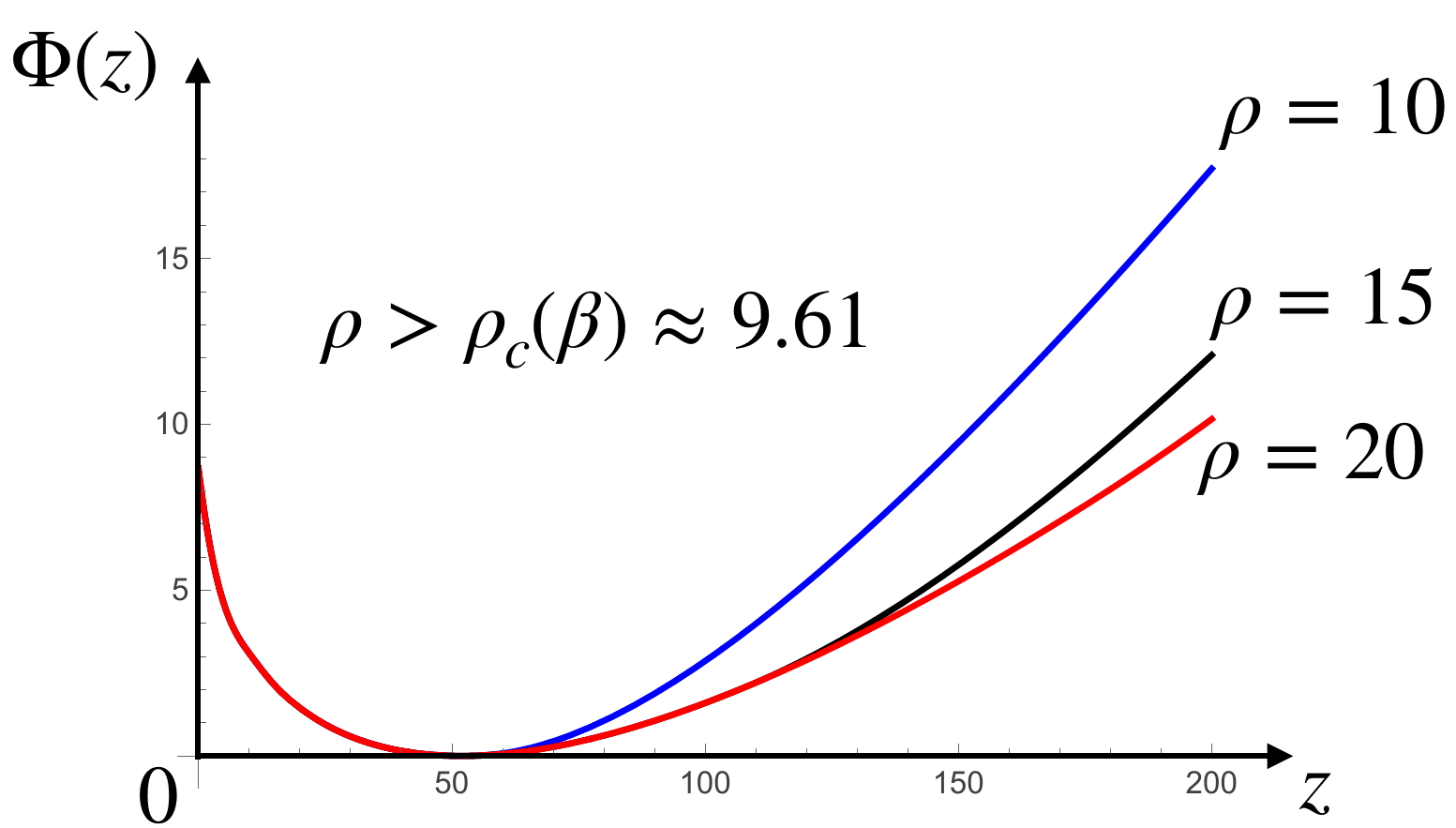}
\caption{\mk{Plot of $\Phi(z)$ vs $z$ for different values of $\rho > \rho_c(\beta) \approx 9.61$ with $\rho = 10, 15$ and $\rho = 20$ from top to bottom. Here $d=3$ and $\beta = 1/2$, in which case $z_c \approx 51.95, 109.1$ and $176.2$ for $\rho = 10, 15$ and $20$ respectively. One sees that the part of the curve for $z<z_c$ becomes frozen, i.e., remains independent of $\rho$ as $\rho$ increases.}}\label{Fig_rhogeqrhoc}
\end{figure}

It is interesting to compare the locations of $z_{\min}$, where $\Phi(z)$ reaches its global minimum, and the critical point 
$z_c$, at which $\Phi(z)$ is nonanalytic. This is best understood in terms of the phase diagram in the $(\lambda, \rho)$ plane 
in Fig.~\ref{Fig_phdiag}. \mk{We fix $\rho$ and decrease $z$ by increasing $\lambda$}. When $z \to z_{\min}$, we have seen that it corresponds to approaching $\lambda=0$ along a 
horizontal axis corresponding to a fixed $\rho$. On the other hand, when $z$ approaches $z_c$, it corresponds to $\lambda$ 
approaching $\lambda_c$ [given in Eq. (\ref{lmax})] along the same horizontal axis. Thus, as long as $\lambda_c >0$, we must have $z_c < z_{\min}$. On the 
other hand, if $\lambda_c<0$, then $z_c > z_{\min}$. From the exact expressions of $\lambda_c$ in Eq. (\ref{lmax}) and 
$\rho_c(\beta)$ from Eq. (\ref{rhoc_tilde}), one sees that $\lambda_c$ becomes $0$ exactly when $\rho = \rho_c(\beta)$. For 
$\rho < \rho_c(\beta)$, one has $\lambda_c >0$ and hence $z_{\min}<z_c$. In contrast, for $\rho>\rho_c(\beta)$, clearly 
$\lambda_c<0$ and, hence, $z_c>z_{\min}$ (see Fig. \ref{Fig_rate} for a plot of $\Phi(z)$ in the two cases
$\rho<\rho_c(\beta)$ and $\rho>\rho_c(\beta)$). Exactly at $\rho = 
\rho_c(\beta)$, where $\lambda_c=0$, the two points $z_{\min}$ and $z_c$ coincide. In the $(\lambda, \rho)$ plane, this 
special critical point corresponding to $\lambda_c=0$ is shown by a red bullet in Fig. \ref{Fig_phdiag}, where, interestingly all the four phases F-F, 
F-C, C-F and C-C meet, indicating that this is a ``multi-critical'' point. The jump of the second \mk{derivative of $\Phi(z)$ 
also vanishes} exactly at this multi-critical point $z=z_c$. 

\mk{Thus, to summarize, the following picture emerges for the distribution of $P_\beta(I,N)$ as one varies the control parameter $\rho$ on the condensed side. First, we recall that the mean [Eq.~\eqref{av_I_final}] and the variance [Eq.~\eqref{varI_final}] of $I$ freeze as $\rho$ increases through $\rho_c(\beta)$, i.e., they become independent of the density.} \mk{However, this freezing phenomenon extends beyond the first two cumulants and in fact to the full large deviation function as we discuss now. Consider first $\rho\to \rho_c(\beta)^+$. From the phase diagram in Fig.~\ref{Fig_phdiag}, we see that we are in the C-C phase if $\lambda>\lambda_c$. Consequently from Eq.~\eqref{Psi_CF_CC} and the second line of Eq.~\eqref{summary}, we get 
\begin{equation}
\label{eq:psi_lam}
\Psi(\lambda) = \frac{\zeta(d+1)}{\beta^{d}} - \frac{\zeta(d+1)}{\big[\beta(\beta+2\lambda)\big]^{\frac{d}{2}}}\,.
\end{equation}
Eq.~\eqref{eq:psi_lam} holds for $\lambda>\lambda_c$ and is explicitly independent of $\rho$. The $\rho$ dependence enters only through the boundary value $\lambda_c$ [Eq.~\eqref{lmax}]. Consequently, the rate function $\Phi(z)$ in Eq.~\eqref{inv_Legendre} also becomes independent of $\rho$ for $z<z_c$ where $z_c$ is given in Eq.~\eqref{zc}}. \mk{As $\rho$ increases further the value of $z_c$, which depends on $\rho$, increases but the functional form of $\Phi(z)$ for $z<z_c$ remains the same (see Fig.~\ref{Fig_rhogeqrhoc}). Thus the signature of the condensation in $P_\beta(I, N)$ is quite unusual in the condensed phase. The part of the distribution for $z<z_c$ do not change as more particles are pumpled in by increasing $\rho$. The added particles change only the form of the distribution for $z>z_c$, resembling a traveling front scenario.}

We end this section with one interesting observation. Consider the original system to be in the fluid phase where $\rho<\rho_c(\beta)$. If 
we stay in this fluid phase and do not change the density or the temperature, we do not have access to the condensed side. In 
other words, the partition function of the system has no information about the condensed side, unless one tunes the density or 
the temperature through the critical line $\rho_c(\beta)=\zeta(d)/\beta^d$. However, if one probes the rate function $\Phi(z)$ 
associated with the distribution of the moment of inertia $I = z\, V$, while staying completely inside the fluid phase, one 
finds the critical point $z_c < z_{\min}$ where the rate function undergoes a second order jump, signaling the 
existence of a condensed phase. In other words, we can set all the parameters of the system completely on the fluid side and 
yet access the signature of the condensed phase by choosing an appropriate observable (such as the moment of inertia here) and 
studying the rate function associated to its distribution. This is somewhat similar in spirit to devising a ``tilted'' 
Hamiltonian in importance sampling algorithms to access large deviation tails that describe rare events not 
captured by the typical statistics associated to a non-tilted Hamiltonian~\cite{Hartmann2002,Bucklew_book,Krauth_book}.

\section{Conclusion}\label{sec:conclusion}

In this paper, we have studied, in all dimensions and at all temperature, the full probability distribution of the moment of 
inertia $I \propto \sum_{i=1}^N \vec{r}_i^{\,2}$ of $N$ noninteracting trapped bosons in a harmonic potential $V(r) = (1/2) m 
\omega^2 r^2$. In the thermodynamic limit $N \to \infty$ and $\omega\to 0$ with the product $\rho = N \omega^d$ fixed (that 
plays a role analogous to the density), and in dimension $d>1$, the system undergoes a BEC phase transition, from a fluid 
phase at low density ($\rho<\rho_c(\beta)$) to a high density condensed phase (when $\rho > \rho_c(\beta)$) where 
$\rho_c(\beta) = \zeta(d)/\beta^d$, with $\beta$ denoting the inverse temperature. 
\mk{Even though the condensation occurs in all $d>1$, close to the transition point when $\rho$ approaches $\rho_c(\beta)$, the analysis turns out to be technically more cumbersome for $1<d\leq 2$ where the condensate is anomalous, while it is much simpler for the normal condensate that occurs for $d>2$. Hence, when analysing the system near the critical point, we restricted ourselves to $d>2$
for simplicity. The detailed analysis near the condensation transition in the anomalous case $1<d\le 2$ and comparisons to numerical simulations for $d=2$ will
be presented elsewhere~\cite{HKMS}.}

We showed, that in the thermodynamic limit, 
the probability distribution $P_\beta(I,N)$ admits a large deviation form $P_\beta(I,N) \sim e^{-V \Phi(I/V)}$ where $V = 
\omega^{-d} \gg 1$, with a nontrivial rate function $\Phi(z)$ that we have computed analytically. The occurrence of the BEC 
transition in $d>1$ introduces a singularity in the rate function $\Phi(z)$ at a critical value $z_c$, where its second 
derivative undergoes a discontinuous jump. \mk{We show that, on the condensed side, $\Phi(z)$ for $z<z_c$ gets frozen (independent of $\rho$) when $\rho$ exceeds $\rho_c(\beta)$, providing a signature of the ``condensate" in this PDF of $I$ on the condensed side.} One of the striking conclusions of our work is that, even if the actual system is 
in the fluid phase, i.e., when $\rho < \rho_c(\beta)$, by measuring the distribution of $I$ and analysing the singularity in 
the associated rate function, one can get a signal of the BEC transition in $d>1$ without having to tune the actual
density through $\rho=\rho_c(\beta)$.

In this paper, we focused on a specific linear statistics, namely $I \propto \sum_{i=1}^N \vec{r}_i^{\,2}$. In general, 
computing the full distribution of $I$ is highly nontrivial even for noninteracting bosons, since one has to consider both 
quantum and thermal fluctuations, arising respectively from the Bose statistics and the finite temperature. However, for 
noninteracting bosons in a harmonic trap, we could perform this computation due to a nontrivial identity which
we call a duality relation, that relates the 
cumulant generating function of $I$ to the canonical partition function $Z(N,\beta)$. {A first natural question is whether one 
can extend our calculations to anisotropic harmonic traps. 
 However, note that the moment of inertia is an isotropic observable. Hence, the basic identity  in Eqs.~\eqref{relation_tilde} and \eqref{rel_btilde} that we heavily used would no longer be applicable for anisotropic traps. 
Another interesting extension would be to non-harmonic traps, where such a duality relation is non-existent.} We expect the associated rate 
function to depend explicitly on the potential. However, the singularity we found here in the rate function should also be 
present for a generic potential, provided it allows a BEC transition in the thermodynamic limit. Proving this is an 
interesting open problem. Furthermore, one can also ask whether such a singularity in the rate function shows up (for 
noninteracting bosons exhibiting a BEC transition) for other linear statistics of the form ${\cal L} = \sum_{i=1}^N f({\vec 
r}_i)$ where $f({\vec r})$ can be any arbitrary smooth function. For a generic $f(\vec r)$, computing the full 
distribution of ${\cal L}$ is highly 
nontrivial, even for noninteracting bosons in a harmonic trap. For the special case where $f({\vec r}) = {\vec r}{\,^2}$, we could 
compute this distribution thanks to the nice duality relation that we derived in this paper. It would be interesting and challenging 
to see if similar duality relations can be derived for generic $f({\vec r})$.

During the last years, there has been impressive numerical advances using importance sampling techniques to measure the rate 
functions associated with a probability distribution, up to very high precision. In particular, singularities in the rate 
function have been numerically detected in a variety of models (see ~\cite{BMR2024,WH2024} for recent references). 
It would be interesting to see if the rate function computed analytically in this paper can be measured numerically. \mk{It would be interesting to investigate numerically how the shape of the distribution
$P_\beta(I,N)$ vs. $I$ changes, as the density $\rho$ increases through the critical value
$\rho_c(\beta)$.
In particular, how does the distribution look exactly at the critical point $\rho=\rho_c(\beta)$ ?
We made some progress in these directions which will be reported elsewhere~\cite{HKMS}.}

The moment of inertia $I$ is a rather natural observable whose statistics can hopefully be measured in cold atom 
experiments. In fact, in typical many-body quantum systems, exact results can sometimes be obtained only at zero temperature $T=0$. 
At finite temperature, where most experimental data are available, very few exact results exist. Our work 
provides an exact finite temperature result for the statistics of $I$ and hence, in principle, it would be possible to compare 
the future experimental data with our theoretical predictions.

{In this work, we considered only noninteracting bosons. It would be interesting to study observables like moment of inertia for weakly interacting bosons. For example, under the Bogoliubov approximation, weakly interacting bosons may effectively be reduced to noninteracting harmonic oscillators and it would be interesting to see if our method can be extended to study the moment of inertia in such a weakly interacting Bose gas.}

\section{Acknowledgments}
\label{sec:ack}

{\mk{We thank A. K. Hartmann for useful discussions}. SNM and GS acknowledge support from ANR Grant No. ANR-23-
CE30-0020-01 EDIPS.
MK acknowledges the support of the Department of Atomic Energy, Government of India, under project no. RTI4001.}

\newpage

\newpage

\appendix

\section{Derivation of the duality relation in $d$ dimensions}\label{app_duality}

In this appendix, we derive the duality relation stated in Eqs. (\ref{relation_tilde}) and (\ref{rel_btilde}).
We consider the Hamiltonian of $N$ noninteracting bosons in a harmonic trap in dimension $d$, namely
\bea \label{H_N_app}
\hat H_N(m) = \sum_{i=1}^N \hat h_i(m) \quad {\rm where} \quad 
\hat h_i(m) = \frac{\vec{p}_i{\,^2}}{2m} + \frac{1}{2} m \omega^2 \vec{r}_i^{\, 2}  \; .
\eea
We want to compute the quantum average $\langle e^{-\lambda I }\rangle$, with $I = \omega^2 \sum_{i=1}^N \vec{r}_i^{\, 2}$. 
For this we need to compute 
\bea \label{Trace1}
{\rm Tr} \left[e^{-\lambda  \omega^2 \sum_{i=1}^N \vec{r}_i^{\, 2}}\, e^{- \beta \hat H_N}  \right] =  
{\rm Tr} \left[e^{-\frac{\lambda}{2}  \omega^2 \sum_{i=1}^N \vec{r}_i^{\,2} }\, e^{- \beta \hat H_N}  
e^{-\frac{\lambda}{2}  \omega^2 \sum_{i=1}^N \vec{r}_i^{\, 2}}\right] = {\rm Tr}\left[ \prod_i^N \hat O_i \right] \;,
\eea
where $\hat O_i = e^{-\frac{\lambda}{2} \vec{r}_i^{\, 2}} e^{-\beta \hat h_i(m)} 
e^{-\frac{\lambda}{2} \vec{r}_i^{\,2}}$. In the first 
equality, we have used the cyclicity of the trace while in the second equality we have used the fact that the bosons are 
noninteracting (and hence their Hamiltonians commute). 
Let us now consider a generic matrix element of the operator $\hat O_i$, namely
\mk{\bea \label{matrix_elem}
\langle \vec{y}_i \vert e^{-\frac{\lambda}{2} \vec{r}_i^{\,2}}\, e^{-\beta \hat{h}_i(m)}\,
 e^{-\frac{\lambda}{2} \vec{r}_i^{\,2}} \vert \vec{y}'_i \rangle = 
e^{-\frac{\lambda \omega^2}{2}(\vec{y}_i^2 + \vec{{y}'}_i^2)} \langle \vec{y}_i \vert e^{-\beta \hat h_i(m)} 
\vert \vec{y}'_i \rangle \, . 
\eea}
The matrix element on the right hand side of Eq. (\ref{matrix_elem}) factorises into $d$ factors, one for each
each component in $d$ dimensions.
Let us now recall the explicit expression of the propagator of the $1d$ quantum harmonic oscillator (in imaginary time), namely
\begin{equation} \label{propag_1d}
\langle y \vert e^{-\beta \hat h(m)} \vert y' \rangle = \sqrt{\frac{m \omega}{2 \pi \hbar \sinh{(\beta \hbar \omega)}}} \exp{\left[- \frac{m \omega}{2 \hbar \tanh{(\beta \hbar \omega)}}(y^2 + {y'}^2) + 
\frac{m \omega}{\hbar \sinh{(\beta \hbar \omega)}}(y\, y') \right]}\;.
\end{equation}
We next observe that the external factor $e^{-\frac{\lambda \omega^2}{2}(\vec{y}_i{^2} + \vec{{y}'}_i^2)}$ on the
right hand side of Eq. (\ref{matrix_elem})
can be absorbed in the quadratic form of the harmonic oscillator propagator in Eq. (\ref{propag_1d}, leading us to 
rewrite Eq. (\ref{matrix_elem}) as
\bea \label{duality}
\langle \vec{y}_i \vert e^{-\frac{\lambda}{2} \vec{r}_i^{\,2}} e^{-\beta \hat{h}_i(m)} 
e^{-\frac{\lambda}{2} \vec{r}_i^{\,2}} \vert \vec{y}^{\,'}_i \rangle = \langle  \vec{y}_i  \vert 
e^{-\tilde \beta \hat h_i(\tilde m)} \vert \vec{y}^{\,'}_i \rangle 
\eea
in terms of renormalised parameters $\tilde m$ and $\tilde \beta$
\bea 
&&\tilde m = m \frac{\sinh(\tilde \beta \hbar \omega)}{\sinh{(\beta \hbar \omega)}} \;, \label{mtilde} \\
&& \cosh{(\tilde \beta \hbar \omega)} = \cosh{(\beta \hbar \omega)}  + \lambda \frac{\hbar \omega}{m} \,  \sinh{(\beta \hbar \omega)} \;.
\eea
\mk{Therefore, using (\ref{Trace1}) together with (\ref{duality}) one has the relation}

\mk{\begin{equation}
\label{eq:inter}
\left\langle
e^{-\lambda \omega^2 \sum_i r_i^2}
\right\rangle
=
\frac{1}{Z(N,\beta)}
\operatorname{Tr}
\left[
e^{-\lambda \omega^2 \sum_i r_i^2 - \beta H_N(m)}
\right]
=
\frac{1}{Z(N,\beta)}
\operatorname{Tr}
\left[
e^{-\tilde{\beta} H_N(\tilde m)}
\right].
\end{equation}}
\mk{Since the spectrum of the Hamiltonian $H_{N}(m)$ does not depend on $m$ explicitly, it follows that}
\mk{\begin{equation}
\operatorname{Tr}\!\left[
e^{-\tilde{\beta} H_N(\tilde{m})}
\right]
=
\operatorname{Tr}\!\left[
e^{-\tilde{\beta} H_N(m)}
\right] \equiv Z(N, \tilde{\beta})
\label{eq:m_ind}
\end{equation}}

\mk{Using Eq.~\eqref{eq:m_ind}, we then get from Eq.~\eqref{eq:inter}, the key relation in terms of ratios of partition functions}
\bea \label{ratio}
\langle e^{- \lambda \omega^2 \sum_{i=1}^N \vec{r}_i^{\,2}} \rangle = \frac{1}{Z(N,\beta)} {\rm Tr} 
\left[e^{-\lambda  \omega^2 \sum_{i=1}^N \vec{r}_i^{\, 2}}\, e^{- \beta \hat H_N}  \right] = \frac{Z(N,\tilde \beta)}{Z(N,\beta)}\,, 
\eea
with
{\bea
Z(N,\beta) = {\rm Tr} \left[e^{-\beta \hat H_N(m)} \right]\, .
\eea}
Note also that this relation gives the Laplace transform of $P(I,N)$ since
\bea
\tilde P(\lambda, N) = \int_0^{\infty} e^{- \lambda I} P(I,N) d I = \langle e^{- \lambda \omega^2 \sum_{i=1}^N 
\vec{r}_i^{\, 2}} \rangle \;.
\eea

\mk{We would like to end this section with an important note. If the spectrum of $H_N(m)$ depended explicitly on the mass $m$, then $\operatorname{Tr}\!\left[
e^{-\tilde{\beta} H_N(\tilde{m})}
\right]$
would represent the partition function of a renormalized Hamiltonian. However, it would not be possible to express it as the function
$Z(N,\tilde{\beta})$, where $Z(N,\beta)$ denotes the partition function of the bare Hamiltonian. The fact that the Laplace transform of $I$ can be expressed as the ratio of the same function $Z(N,\beta)$ evaluated at different values of $\beta$ depends crucially on the fact that the spectrum of the harmonic oscillator does not depend explicitly on the mass $m$.}

\section{Computation of the mean and the variance of $I$}\label{App_var}

We first recall the relation 
\bea  \label{rel_btilde_appvar}
 \cosh{(\tilde \beta(\lambda) \hbar \omega)} = \cosh{(\beta \hbar \omega)}  + \lambda \,{\hbar \omega} \,  \sinh{(\beta \hbar \omega)} \;.
\eea
Taking derivative with respect to $\lambda$ one gets
\bea \label{dbtilde}
\frac{\partial \tilde \beta(\lambda)}{\partial \lambda} =  \frac{\sinh{\left(\beta \hbar \omega\right)}}{\sinh{\left(\tilde \beta(\lambda) \hbar \omega\right)}} \;.
\eea
Taking one more derivative with respect to $\lambda$ and using the chain rule $\partial_\lambda = (\partial \tilde \beta/\partial \lambda) \, \partial_{\tilde \beta}$ gives
\bea \label{dbtilde2}
\frac{\partial^2 \tilde \beta}{\partial \lambda^2} = - \hbar \omega \, {\rm coth}\left(\tilde \beta(\lambda) \hbar \omega \right) \frac{\sinh^2{(\beta \hbar \omega)}}{\sinh^2{(\tilde \beta(\lambda) \hbar \omega)}}\;.
\eea
Next we recall from Eq. (\ref{relation_tilde}) that the cumulant generating function of $I$ is given by 
\bea \label{relation_tilde_app}
\langle e^{-\lambda I}\rangle = \frac{Z(N,\tilde \beta(\lambda))}{Z(N,\beta)} \;.
\eea 
Taking derivative with respect to $\lambda$ and setting $\lambda=0$ gives the expression for $\langle I \rangle$
\bea \label{av_I_app1}
\langle I \rangle= - \frac{\partial}{\partial \lambda}
\langle e^{-\lambda I}\rangle \Big \vert_{\lambda = 0} = - \frac{\partial}{ \partial \tilde \beta} \left[\frac{Z(N,\tilde \beta(\lambda))}{Z(N,\beta)}\right] \, \frac{\partial \tilde \beta(\lambda)}{\partial \lambda} \Bigg \vert_{\lambda=0}
\eea
Using the relation in Eq. (\ref{dbtilde}) for $\lambda = 0$ and $\tilde \beta(\lambda=0)= \beta$ then gives 
\bea \label{av_I_app2}
\langle I \rangle = - \frac{\partial \ln Z(N,\beta)}{\partial \beta} \;.
\eea

We now compute the variance ${\rm Var}(I)$ which can be obtained from the cumulant generating function as
\bea \label{varI_app_CGF}
{\rm Var}(I) = \frac{\partial^2}{\partial \lambda^2} \ln \langle e^{-\lambda I}\rangle \Big \vert_{\lambda = 0} \;.
\eea
Using the relation in Eq. (\ref{relation_tilde_app}) one finds
\bea\label{varI_app_CGF2}
{\rm Var}(I) = \frac{\partial^2}{\partial \lambda^2} \left[ \ln Z(N,\tilde \beta(\lambda))\right] \Big \vert_{\lambda=0} \;.
\eea
Using again the chain rule $\partial_\lambda = (\partial \tilde \beta/\partial \lambda) \, \partial_{\tilde \beta}$ we then find
\bea\label{varI_app_CGF3}
{\rm Var}(I) = \frac{\partial^2}{\partial \tilde \beta^2} \left[\ln Z(N,\tilde \beta)\right] \,
\left( \frac{\partial \tilde \beta}{\partial \lambda}\right)^2 \Bigg \vert_{\lambda=0} + 
\frac{\partial}{\partial \tilde \beta} \left[\ln Z(N,\tilde \beta)\right] \frac{\partial^2 \tilde \beta}{\partial \lambda^2} \Bigg \vert_{\lambda=0}
\eea
Putting together the results from Eqs. (\ref{dbtilde}), (\ref{dbtilde2}) and (\ref{av_I_app2}) one gets
\bea \label{rel_var_app}
{\rm Var}(I) = \langle I^2 \rangle -  \langle I \rangle^2 = \left[ \frac{\partial^2 \ln Z}{\partial \beta^2} + {\hbar \omega}\, {\rm coth}(\beta \hbar \omega) \langle I \rangle\right] \;.
\eea

\section{The computation of a multidimensional integral}
\label{App_kintegral}

In this Appendix we compute the integral in Eq. (\ref{sp_eq}), namely
\bea \label{sp_eq_app}
\rho = \frac{1}{\beta^d}\int_> d\vec k \frac{1}{e^{s^*+\sum_{i=1}^d k_i}-1} \;.
\eea
We first rewrite the integral as
\bea\label{sp_eq_app2}
\rho = \frac{1}{\beta^d}\int_0^\infty dq \int_> d\vec k  \frac{1}{e^{s^*+q-1}} \delta\left(q-\sum_{i=1}^d k_i \right) \;.
\eea
We then perform first the integral over ${\vec k}$ as follows. We define
\bea \label{def_Iofq}
I(q) = \int_> d{\vec k} \, \delta \left(q-\sum_{i=1}^d k_i \right) \;.
\eea
Taking the Laplace transform with respect to $q$ decouples the integral into $d$ separate one-dimensional integrals, leading to
\bea
\int_0^\infty e^{-sq}\, I(q) dq = \frac{1}{s^d}\;.
\eea 
Inverting the Laplace transform trivially we get
\bea \label{Iofq2}
I(q) = \frac{1}{\Gamma(d)}q^{d-1} \;.
\eea
{Substituting Eq. (\ref{Iofq2}) in Eq. (\ref{sp_eq_app2}) gives}
\bea \label{sp_eq_app3}
\tilde \rho = \rho \, \beta^d = \frac{1}{\Gamma(d)} \, \int_0^{\infty} dq\, q^{d-1}  \frac{1}{e^{s^*+q}-1}\, .
\eea
Now, expanding the integrand in powers of $e^{-q}$ and integrating term by term gives the result in Eq. (\ref{sp_eq2}).

\end{document}